%% file: 0paper.tex
\documentclass[sigconf,screen,nonacm]{acmart}

\input{00packages.tex}

\AtBeginDocument{%
  \providecommand\BibTeX{{%
    \normalfont B\kern-0.5em{\scshape i\kern-0.25em b}\kern-0.8em\TeX}}}

\begin{document}

\title{Detecting Early and Implicit Suicidal Ideation via Longitudinal and Information Environment Signals on Social Media}



\author{Soorya Ram Shimgekar}
\orcid{0000-0002-1110-9699}
\affiliation{%
 \institution{University of Illinois Urbana-Champaign}
 \city{Urbana}
 \state{IL}
 \country{USA}}
 \email{sooryas2@illinois.edu}

\author{Ruining Zhao}
\orcid{0009-0009-1700-9101}
\affiliation{%
 \institution{University of Illinois Urbana-Champaign}
 \city{Urbana}
 \state{IL}
 \country{USA}}
 \email{ruining9@illinois.edu}

 \author{Agam Goyal}
\orcid{0009-0009-5989-2887}
\affiliation{%
 \institution{University of Illinois Urbana-Champaign}
 \city{Urbana}
 \state{IL}
 \country{USA}}
 \email{agamg2@illinois.edu}

 \author{Violeta J. Rodriguez}
\orcid{0000-0001-8543-2061}
\affiliation{%
 \institution{University of Illinois Urbana-Champaign}
 \city{Urbana}
 \state{IL}
 \country{USA}}
 \email{vjrodrig@illinois.edu}

 \author{Paul A. Bloom}
\orcid{0000-0003-3970-5721}
\affiliation{%
 \institution{Columbia Irving Medical Center}
 \institution{New York State Psychiatric Institute}
 \city{New York}
 \state{New York}
 \country{USA}}
 \email{paul.bloom@nyspi.columbia.edu}

 \author{Navin Kumar}
\orcid{0000-0003-4502-069X}
\affiliation{%
 \institution{Nimblemind}
 \city{New York City}
 \state{New York}
 \country{USA}}
 \email{navin@nimblemind.ai}

\author{Hari Sundaram}
\orcid{0000-0003-3315-6055}
\affiliation{%
 \institution{University of Illinois Urbana-Champaign}
 \city{Urbana}
 \state{IL}
 \country{USA}}
 \email{hs1@illinois.edu}

\author{Koustuv Saha}
\orcid{0000-0002-8872-2934}
\affiliation{%
 \institution{University of Illinois Urbana-Champaign}
 \city{Urbana}
 \state{IL}
 \country{USA}}
 \email{ksaha2@illinois.edu}

\renewcommand{\shortauthors}{Shimgekar et al.}

\input{0abstract}



\maketitle


\input{1introduction_new.tex} 
\input{2relatedwork.tex} 
\input{3data.tex} 
\input{4methods} 
\input{5results} 
\input{6discussion} 
\input{7conclusion.tex} 


\begin{acks}

Shimgekar and Saha were partly supported by a gift award from Nimblemind.

\end{acks}

\balance
\bibliographystyle{ACM-Reference-Format}
\bibliography{0paper}

%
\end{document}
\endinput



%% file: 00packages.tex
\usepackage{color, colortbl, xcolor}
\usepackage{url}
\usepackage{subcaption}
\usepackage{textcomp}
\usepackage{soul}
\usepackage{multirow}
\usepackage{enumitem}
\usepackage{mathtools}
\usepackage{siunitx}
\usepackage{tabularx}

\usepackage{booktabs} 
\usepackage{longtable,booktabs}

\usepackage{multicol}
\usepackage{natbib}  

\usepackage{arydshln}
\definecolor{linkColor}{RGB}{6,125,233}
\definecolor{green}{rgb}{0.0, 0.65, 0.31}
\definecolor{bleudefrance}{rgb}{0.19, 0.55, 0.91}
\definecolor{ceruleanblue}{rgb}{0.16, 0.32, 0.75}
\definecolor{grey}{HTML}{969696}
\definecolor{violet}{HTML}{756bb1}
\definecolor{dgrey}{HTML}{01665e}
\definecolor{lgrey}{HTML}{5ab4ac}
\definecolor{dgreen}{HTML}{005a32}
\definecolor{purple}{HTML}{ae017e}
\definecolor{soothinggreen}{HTML}{C1FFC1}

\definecolor{editCol}{HTML}{0000FF}
\definecolor{maskCol}{HTML}{c51b7d}
\definecolor{lrColor}{HTML}{8856a7}
\definecolor{trColor}{HTML}{d01c8b}
\definecolor{ctColor}{HTML}{4dac26}
\definecolor{brickred}{HTML}{f03b20}

\definecolor{improveCol}{HTML}{4dac26}
\definecolor{worsenCol}{HTML}{d01c8b}

\definecolor{DarkBlue}{HTML}{00008B}
\definecolor{mscolor}{HTML}{01665e}
\definecolor{nmscolor}{HTML}{bf812d}
\definecolor{lgreen}{HTML}{ccece6}
\definecolor{dolive}{HTML}{308014}

\colorlet{tablerowcolor4}{gray!50} 

\newcommand*{\textlabel}[2]{%
  \edef\@currentlabel{#1}
  \phantomsection
  #1\label{#2}
}

\colorlet{tableheadcolor}{gray!25} 
\colorlet{tablerowcolor}{gray!10} 
\colorlet{tablerowcolor2}{gray!45} 
\colorlet{tablerowcolor3}{gray!12} 
\colorlet{tablerowgreen}{soothinggreen!75} 

\newcommand{\rowcollight}{\rowcolor{tablerowcolor3}} %
\newcommand{\rowcolgreen}{\rowcolor{tablerowgreen}} %

\newcolumntype{a}{>{\columncolor{tablerowcolor}}r}
\definecolor{aicolor}{HTML}{018571}
\definecolor{occolor}{HTML}{ff7799}

\definecolor{aicolor}{HTML}{fc8d62}
\definecolor{occolor}{HTML}{253494}

\newif{\ifhidecomments}
  \hidecommentsfalse 
\ifhidecomments
    \newcommand{\soorya}[1]{}
    \newcommand{\ruining}[1]{}
    \newcommand{\violeta}[1]{}
    \newcommand{\hari}[1]{}
    \newcommand{\koustuv}[1]{}
\else
    \newcommand{\soorya}[1]{\textbf{\small\sffamily{\textcolor{DarkBlue}{[#1 -- Soorya]}}}}
    \newcommand{\ruining}[1]{\textbf{\small\sffamily{\textcolor{orange}{[#1 -- Ruining]}}}}
    \newcommand{\violeta}[1]{\textbf{\small\sffamily{\textcolor{violet}{[#1 -- Violeta]}}}}
    \newcommand{\hari}[1]{\textbf{\small\sffamily{\textcolor{dgreen}{[#1 -- Hari]}}}}
    \newcommand{\koustuv}[1]{\textbf{\small\sffamily{\textcolor{purple}{[#1 -- Koustuv]}}}}
  \fi

\newcommand{\Cs}{\textit{Case}}
\newcommand{\Ct}{\textit{Control}}

\renewcommand{\textrightarrow}{$\rightarrow$}

\newcommand{\sitb}{SI}
\newcommand{\rsw}{\textit{r/SuicideWatch}}

\colorlet{tableheadcolor}{gray!25} 

\definecolor{neutralCol}{HTML}{dd1c77}
\definecolor{neutralGreen}{HTML}{31a354}
\definecolor{NewBlue}{HTML}{1879ba}
\definecolor{bleudefrance}{rgb}{0.19, 0.55, 0.91}  
\definecolor{AfTrColor}{HTML}{0868ac}  
\definecolor{BfTrColor}{HTML}{a8ddb5}  

\definecolor{AfCtColor}{HTML}{b10026}  
\definecolor{BfCtColor}{HTML}{fd8d3c}

\graphicspath{ {figures/} }

\newcommand{\para}[1]{\vspace{0.2em}\noindent\textbf{\textit{#1}~}}

\usepackage{booktabs}
\usepackage{tabularx}
\usepackage{lipsum} 
\usepackage{multirow}

%% file: 0abstract.tex
\begin{abstract}

On social media, several individuals experiencing suicidal ideation (SI) do not disclose their distress explicitly. 
Instead, signs may surface indirectly through everyday posts or peer interactions. 
Detecting such implicit signals early is critical but remains challenging.
We frame early and implicit SI as a forward-looking prediction task and develop a computational framework that models a user's information environment, consisting of both their longitudinal posting histories as well as the discourse of their socially proximal peers. 
We adopted a composite network centrality measure to identify top neighbors of a user, and temporally aligned the user's and neighbors' interactions---integrating the multi-layered signals in a fine-tuned DeBERTa-v3 model.
In a Reddit study of 1,000 (500 \Cs{} and 500 \Ct{}) users, our approach improves early and implicit SI detection by an average of 10\% over all other baselines. These findings highlight that peer interactions offer valuable predictive signals and carry broader implications for designing early detection systems that capture indirect as well as masked expressions of risk in online environments.
\end{abstract}

%% file: 1introduction_new.tex
\section{Introduction}
More than 720,000 individuals die by suicide each year worldwide, making it a major global public health concern~\cite{world2025suicide}. 
Suicidal ideation (\sitb{}) refers to a spectrum of cognitive and behavioral manifestations related to suicide, ranging from passive thoughts of death to active planning and engagement in self-injurious actions with intent to die~\cite{joiner2005interpersonal}. 
While early detection of these signs is critical, psychiatry and mental health services have long struggled to detect risk before individuals disclose it explicitly~\cite{calear2019suicidal}. A recent meta-analysis revealed that the estimated prevalence of \sitb{} disclosure is only 46.3\%, inferring that the majority of people who experience suicidal thoughts do not disclose them, making early detection critical for timely intervention~\cite{hallford2023disclosure}.

Online platforms have become vital spaces where signs of \sitb{} may surface~\cite{shimgekar2025interpersonal, nesi2021social,de2016discovering,de2017language,saha2019social}. 
People use online mental health communities to share distress, seek help, and connect with peers~\cite{de2014mental}. 
This has created opportunities for both researchers and clinicians to understand suicide risk through language, social interactions, and online behavioral cues.
Prior work has shown the value of linguistic cues for identifying mental health risks~\cite{coppersmith2014quantifying,guntuku2017detecting}. 
In particular, prior work has computationally modeled the language of \sitb{} on social media~\cite{alghazzawi2025explainable,zhang2025ketch,saha2019social}.
More recent computational approaches leveraged longitudinal and multimodal data, as well as social network analyses, to anticipate suicide-related behaviors~\cite{shen2020suicide}.

Yet, critical limitations persist. 
Prior research has primarily examined posts where \sitb{} is explicitly disclosed, such as direct or indirect references to self-harm or suicidal thoughts within suicide-related forums or discussions~\cite{ji2022towards,bloomidentifying,saha2019social,burnap2015machine}. 
These approaches presuppose that individuals articulate distress.
However, many at-risk individuals neither disclose \sitb{} nor exhibit overt warning signs, but instead mask distress within seemingly ordinary discourse~\cite{mcgillivray2022non, podlogar2022past}.
Our work targets the detection of these undisclosed \sitb{}, characterized by subtle, contextually obscured, or socially distributed indicators, to potentially enable earlier and more effective intervention.

To address the above gap, our work is guided by the research question (RQ): \textbf{How can early signals of \sitb{} be detected in social media activity, particularly in the absence of explicit disclosures?} 
To address the RQ, we conceptualize early and implicit \sitb{} as a forward-looking prediction problem that requires modeling both an individual's longitudinal behavior and the broader social context in which that behavior unfolds.  
We develop a framework that jointly captures temporal dynamics in a user's posting history and the conversations of socially proximal peers, enabling a richer representation of early warning signals.
This idea of using socially proximal peers to understand an individual's mental state is derived from many prior acclaimed psychology works~\cite{copeland2021long}.
Within this framework, we target three aims:

\para{Aim 1 (Immediate Interaction):} Examine whether users' longitudinal posting patterns, in conjunction with others' comments on their posts, reveal early signals of implicit \sitb{}. 

\para{Aim 2 (Neighbor Interaction):} Examine how users' information environment, including peer interactions and exposure to content, contributes as predictive features for implicit \sitb{}. 

\para{Aim 3:} Identify the linguistic markers that underlie interaction types, in terms of how specific language cues are associated with these interactions.

We conducted our study on Reddit~\cite{saha2020causal, shimgekar2025interpersonal}, focusing on a sample of 1,000 users divided into 500 \Cs{} and 500 \Ct{} users (who never participated in mental health conversations). 
Our predictive framework jointly modeled each user's full posting timeline along with the discourse of their most socially proximal neighbors, identified through network centrality, and fine-tuned a DeBERTa-v3 model~\cite{he2021debertav3} to embed both individual and peer interaction signals into a unified representation.

Our findings show that incorporating peer interactions within the information environment, alongside users' longitudinal posting histories, significantly enhances detection of early and implicit \sitb{}, improving performance by an average of \textbf{10\%} compared to other baseline models based only on longitudinal user data. 
Overall, this paper contributes: 1) a framework for detecting implicit \sitb{}; 2) a method for systematically integrating environment information through neighbor interactions; and 3) empirical evidence that this integration improves early detection.

\para{Privacy, Ethics, and Reflexivity.}
This paper used publicly accessible social media discussions on Reddit and did not require direct interactions with individuals, thereby not requiring ethics board approval. 
However, we are committed to the ethics of the research and followed practices to secure the privacy of individuals in our dataset. 
This paper only presents paraphrased quotes to reduce traceability, yet provides context to readership. 
Our research team comprises researchers holding diverse gender, racial, and cultural backgrounds, including people of color and immigrants, and hold interdisciplinary research expertise. Our research team comprises computer scientists with expertise in HCI, CSCW, and social computing, and psychologists with expertise in clinical psychology, adolescent depression and suicide, and digital health interventions.
One of our psychologist coauthors specializes in suicide etiology, suicide prevention, and crisis intervention, and the other psychologist coauthor is a clinical psychologist with over 16 years of experience spanning adult and adolescent inpatient care and crisis suicide helplines.
To ensure validity and prevent misrepresentation, our findings were reviewed and corroborated by our psychologist coauthors. 
We further discuss the ethical implications of our work in~\autoref{sec:practical_implications}.

%% file: 2relatedwork.tex
\section{Related Work}\label{section:rw}

\subsection{Suicidal Ideation and Social Context.}
Psychological theory views suicidal ideation (\sitb{}) as a consequence of intertwined social and cognitive factors. The Interpersonal Theory of Suicide~\cite{joiner2005interpersonal} posits that suicidal desire arises from perceived burdensomeness and thwarted belongingness, while capability develops through repeated exposure to pain or fear. 
Prior research conducted computational studies of depression and \sitb{} focused on overt signals such as self-disclosure or negative sentiment~\cite{coppersmith2014quantifying, dechoudhury2013predicting,saha2019social}. 
Research revealed that many at-risk individuals express distress through subtle cues, motivating methods that model temporal and semantic behavioral dynamics~\cite{guntuku2017detecting,benton2017multi}.

A range of methods have been proposed to detect mental health risk signals beyond surface cues. 
For instance,~\citet{fatima2021dasentimental} introduced \textit{DASentimental}, a semi-supervised model combining bag-of-words and semantic networks, while~\citet{trotzek2018utilizing} showed that convolutional networks with linguistic metadata enable earlier detection. More recent work leverages large language models, with GPT-3.5/4 using chain-of-thought prompting on diaries~\cite{shin2024using} and reasoning-guided LLMs improving interpretability~\cite{teng2025enhancing}. Temporal dynamics remain critical, from emotional “phases” in user timelines~\cite{sawhney2021phase} to transformer-based models enriched with temporal signals~\cite{sawhney2020statenet}. Social context is equally important: hyperbolic embeddings of user histories and peer interactions enhance prediction~\cite{sawhney2021hyperbolic}, peer networks and conversational responses influence trajectories~\cite{wyman2019peer, de2017language}, and longitudinal patterns reveal precursors of \sitb{}~\cite{de2016discovering}. Complementary directions include clinical-domain datasets such as ScAN~\cite{rawat2022scan}, automated counseling support with PsyGUARD~\cite{qiu-etal-2024-psyguard}, and calls to model underlying intent rather than surface disclosure~\cite{ji2022towards}.

Together, this work provides evidence that \sitb{} arises from psychological distress, temporal dynamics, and social context, demanding models that go beyond surface cues. 
Our study conducts a joint modeling of users’ longitudinal histories and post-level commentary, enabling early detection of implicit \sitb{}. 


\begin{figure*}[t]  
    \centering
    \includegraphics[width=\textwidth]{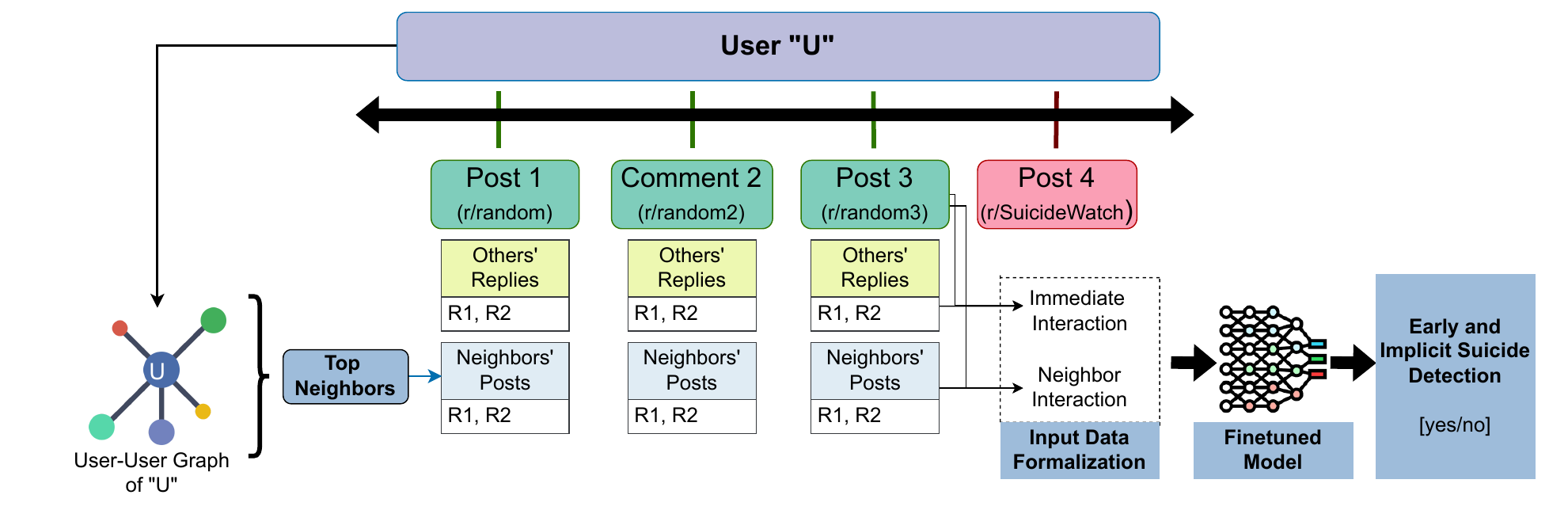} 
    \caption{Illustration of user interactions (Immediate/Neighbor): Immediate interactions include users’ self-posts, self-comments, and received replies, while neighbor interactions include top neighbor posts. Top neighbors identified via \texttt{NeighborScore}.}
    \label{fig:comparison}
\end{figure*}

\subsection{Social Media and Mental Health}
Online platforms are key venues for mental health self-disclosure with communities for peer support and a sense of belonging~\cite{saha2020causal,shimgekar2025interpersonal,de2014mental,zhou2022veteran,andalibi2016understanding,guntuku2017detecting,ernala2017linguistic}.  
Moderated peer-support spaces reduce isolation and help people discuss stigmatized experiences~\cite{johnson2022s,de2014mental,saha2020omhc}. Social support, in the form of both emotional and informational support, has been shown to improve mental wellbeing~\cite{de2017language,saha2020causal,sharma2018mental}.  
Language plays a central role: psycholinguistic research links specific linguistic markers to mental-health outcomes~\cite{pennebaker2007expressive}, and computational studies have used these cues to detect distress and model support dynamics~\cite{guntuku2017detecting,chancellor2020methods,kim2023supporters,saha2020psychosocial}. Prior work has also established the construct validity of these measurements~\cite{saha2022social}.

Recent work has focused on interpretable and fine-grained modeling of mental health on social media. Symptom-based approaches such as \textsc{PsySym}~\cite{zhang2022psysym, chen2023two_stream_mdd} link online language to clinically meaningful categories of disorders. Depression severity has been quantified through semantic similarity to symptom descriptors~\cite{perez2023semantic_severity}, while large language models now enable explainable detection with interpretable rationales~\cite{wang2024_explainable_depression}. Analyses of pre- and post- diagnosis language shifts, highlight the temporal dynamics of distress expression~\cite{alhamed2024_before_after}.
Supportive language marked by adaptability, immediacy, and emotionality predicts better outcomes~\cite{althoff2016large,saha2020causal}, and automatic empathy-detection models scale such insights to peer-support settings~\cite{sharma2020computational}. For \sitb{}, machine-learning approaches have identified risk signals in social media language alongside emotional patterns that precede suicide attempts~\cite{coppersmith2016exploratory,de2016discovering,burnap2015machine,thieme2020machine,yuan2023mental}.

While online data shows promise for early risk detection, most methods isolate either individual language or specific interactions. Our approach instead models full posting timelines alongside peer influences, capturing risk even without explicit \sitb{} disclosures. Unlike prior work that assumes some suicide-related expression in specialized forums, our framework identifies early indicators without requiring references to self-harm or participation in such spaces.

%% file: 3data.tex
\section{Data}\label{sec:data}

We used data from \rsw{}, a semi-anonymous Reddit community focused on \sitb{}, alongside posts and comments from other subreddits. From the PushShift archive~\cite{baumgartner2020pushshift} of April 2019 (18.3M posts, 138.5M comments), the data includes 10,037 posts and 38,130 comments from \rsw{}. Prior work has leveraged Reddit for \sitb{}~\cite{de2017language,de2016discovering,shimgekar2025interpersonal} and mental health studies~\cite{sharma2018mental,saha2022social,saha2020omhc}.

\para{Constructing \Cs{} and \Ct{} datasets.} We identified two user cohorts. The first cohort comprises 500 \Cs{} individuals, defined as individuals who have made at least one post on \rsw{}. 
The second cohort consists of 500 \Ct{} individuals, who never participated in any subreddit related to mental health. 
We identified the \Ct{} users by referencing the taxonomy of mental health-related subreddits from prior work~\cite{sharma2018mental}.

For the \Cs{} group, all posts and comments before each user’s first \rsw{} disclosure were labeled positive (1). On average, users made \textbf{10.07 posts} before disclosure, transitioning within \textbf{1.5 days} of their last non-\rsw{} post. To construct a balanced \Ct{} group, we sampled the first \textbf{10 posts} from each user (matching \Cs{} averages), labeling them negative (0). The dataset was then split 75:25 by users into train and test sets, ensuring no overlap.

For each \Cs{} and \Ct{} user, neighbors are selected solely based on their interaction patterns before their first \rsw{} post, i.e., before any suicidal ideation was explicitly disclosed. Neighbors are not filtered by topics or content, and no suicide-related keywords or signals are used in selection.

\begin{figure}[t]  
    \centering
    \includegraphics[width=0.4\textwidth]{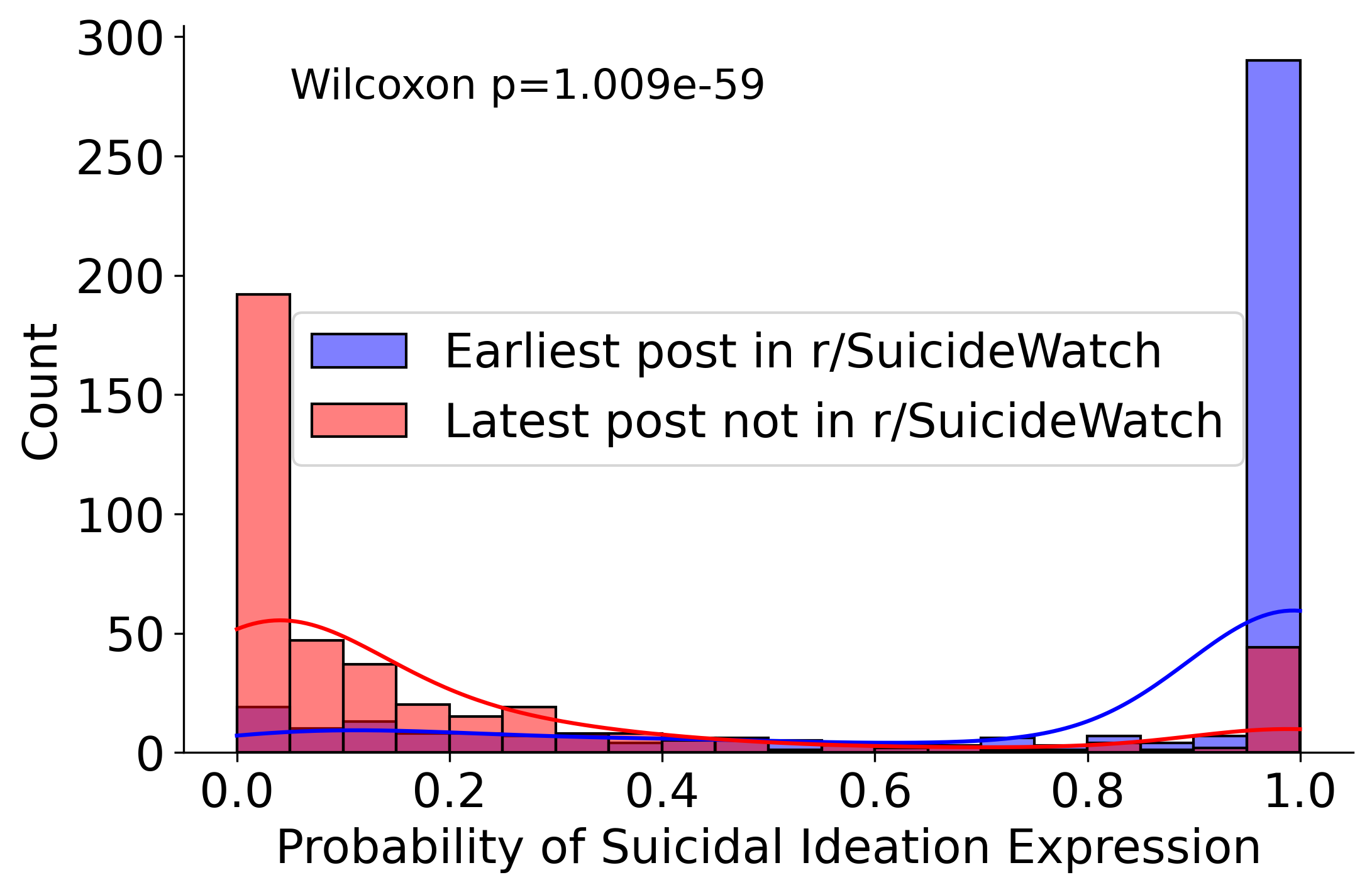} 
     \caption{Distribution of \sitb{} probability for the same set of users before their last non suicidal post and their first post in \texttt{r/SuicideWatch}}
     \label{fig:rsw}
\end{figure}

Our validation reveals that the first post in \rsw{} marks a critical point where implicit ideation becomes explicit disclosure. A zero-shot NLI model~\cite{laurer_less_2023} based on DeBERTa-v3-base, trained on 1.3M hypothesis-premise pairs from eight NLI datasets (e.g., MultiNLI, FEVER-NLI, LingNLI, DocNLI), was used to capture long-range reasoning and assign \sitb{} probabilities to posts. 
Comparing the last non-\rsw{} post with the first \textit{r/Suicide Watch} post reveals a sharp and statistically significant increase in \sitb{} as per the Wilcoxon signed-rank test ($W$ = 3311.000, $p$<0.001), confirming that initial disclosure occurs at the first \rsw{} post (\autoref{fig:rsw}). 

To further validate our assumptions regarding the timing and nature of suicidal ideation (SI) disclosure on Reddit, we conducted a detailed manual review of 150 \Cs{} users’ posting histories. This review was performed by two psychologist co-authors with extensive clinical expertise.
We found that 87\% of \Cs{} users exhibited no explicit suicidal content in their pre-\rsw{} posts, and the remaining 13\% only displayed indirect or vague expressions of distress rather than suicidal ideation. In contrast, 84\% of users' first posts in \rsw{} contained explicit self-disclosures related to suicidal thoughts, self-harm, or acute personal crisis. These findings are consistent with \rsw{}’s posting guidelines, which encourage users to describe their own experiences and direct explicit ideation to this subreddit, while other mental-health subreddits (e.g., r/depression, r/anxiety, r/offmychest) systematically filter out explicit suicidal disclosures through moderation policies~\cite{dechoudhury2016characterizing,de2017language}.

Our psychologist co-authors applied a hierarchical annotation framework inspired by the Interpersonal Theory of Suicide (IPTS)~\cite{joiner2005interpersonal}, categorizing posts along IPTS dimensions (e.g., loneliness, lack of love, self-hate, perceived burdensomeness) and risk factors (thwarted belonging, perceived burdensomeness, acquired capability of suicide), culminating in an assessment of lethally suicidal ideation. Posts exhibiting indicators of lethally suicidal ideation or acquired capability were considered to contain explicit suicidal thoughts and self-harm themes.

%% file: 4methods.tex
\section{Methods}

\subsection{Modeling Early and Implicit \sitb{}}

Towards \textbf{Aim~1} and \textbf{Aim~2}, we study how linguistic signals from a user’s own activity and their surrounding social context contribute to the early detection of implicit \sitb{} risk. 
We frame implicit \sitb{} detection as a forward-looking prediction problem, estimating the likelihood that a user will eventually make a \sitb{} disclosure, proxied by their first post on \rsw{}. 
To capture both active expression and passive exposure, we model user behavior through two interaction types: \textbf{immediate interactions} (self-posts, self-comments, and received comments) and \textbf{neighbor interactions} (posts and comments from top neighbors). Our methodological framework consists of three components: (1) \textit{formalizing interaction data for modeling}, (2) \textit{detecting top neighbors}, and (3) \textit{modeling implicit \sitb{} signals}. Within this framework, we evaluate four state-of-the-art language models—BERT-base-uncased, RoBERTa-large, ELECTRA-large, and DeBERTa-v3-large. An overview of the framework is shown in \autoref{fig:comparison}.

\subsubsection{Formalizing Interaction Data for Modeling}
Towards our Aim 1 and Aim 2, we structured the input to language models to capture user $U$’s timeline, others' replies, and neighbor interactions. 

\para{Immediate Interaction.}  We first aggregated content from a user's timeline, along with others' replies, and then used all of this textual content for fine-tuning the language model.
To train our models, we used different combinations of features---1) $U$'s self-posts, 2) $U$'s self-posts and self-comments, and 3) $U$'s self-posts, self-comments, and others' replies to $U$'s posts or comments. 

\para{Neighbor Interaction.} Neighbor Interaction captures signals from proximate peers.
For this purpose, we temporally aligned the timelines of $U$ and their top-n neighbors. 
Then, at each timestamp $i$, we selected ten posts by the top neighbors closest in time to $U$'s post. 
We aggregated the neighbors' posts with $U$'s posts and embedded them into a dense vector representation. 
This approach led to the fourth type of model, where we included features from both immediate interactions above, as well as the top neighbors' posts.

\subsubsection{Detecting Top Neighbors}
Towards our Aim 2 to understand the impact of neighbor context, for a user $U$, we found their top neighbors by the following steps:

\para{Step 1: Initial User Collection}  
We first collected all users who interacted with the user $U$, defined as either $U$ commenting on their posts/comments or them commenting on $U$'s posts/comments. This neighbor-identification procedure was then applied recursively to a maximum depth $d$=3 and having a minimum of 5 neighbors, ensuring that both direct and indirect neighbors of $U$ were captured.

\para{Step 2: User-User Graph}
Based on all the initial collected users, we constructed a user-user graph where each \textbf{Node} is an individual user, and an undirected \textbf{Edge} connects two users if either of the users has commented on the other's post. The weight of the edge was quantified as the total number of comment-based interactions, with higher weights indicating stronger ties.

\para{Step 3: Top Neighbor Detection}  
From the user-user graph, we identified the top-$n$ ($n$=10) neighbors for $U$. We ranked the neighbors using a \texttt{NeighborScore} ($S(n)$), based on a combined centrality measure, computed as:

\noindent\begin{minipage}{\linewidth}
\footnotesize
\[
\begin{aligned}
S(n) &= C_{\text{in-degree}}(n) + C_{\text{out-degree}}(n) + C_{\text{closeness}}(n) + C_{\text{eigenvector}}(n) \\ &\quad + C_{\text{betweenness}}(n) + C_{\text{PageRank}}(n)
\end{aligned}
\]
\vspace{0em}
\end{minipage}

\noindent$C_{\text{in-degree}}$ and $C_{\text{out-degree}}$ capture normalized connectivity, $C_{\text{betweenness}}$ measures shortest-path centrality, $C_{\text{closeness}}$ denotes proximity, 

\noindent$C_{\text{eigenvector}}$ reflects influence via important neighbors, and $C_{\text{PageRank}}$ estimates probabilistic importance.
The aggregated \texttt{NeighborScore} identifies peers with strong direct ties and broader network influence around $U$. Each centrality metric captures a different facet of network prominence: in-degree, out-degree, closeness, eigenvector centrality, betweenness centrality, and PageRank. While some centralities are correlated, $S(n)$ approximates overall neighbor prominence rather than precise influence. Our choice to combine multiple centrality measures into a single \texttt{NeighborScore} was a pragmatic heuristic rather than a theoretically fixed design. 

\subsubsection{Modeling Implicit \sitb{} Signals}

We framed the task of detecting implicit \sitb{} as a binary classification problem, where each input $x_i$ was labeled $y_i \in {0,1}$ to indicate the absence or presence of risk. Texts were tokenized into subword embeddings, with attention masks applied and sequences padded or truncated to 512 tokens. Each input was then encoded to produce a pooled \texttt{[CLS]} vector of dimension 1024, which was passed through a linear layer to generate logits. Class probabilities were obtained using a softmax function, and the model was optimized with cross-entropy loss using the AdamW optimizer (learning rate $2\times10^{-5}$, weight decay $0.01$) for 20 epochs with a batch size of 8. Performance was evaluated after each epoch on a held-out validation set using accuracy, precision, recall, and F1-score.

Across the four model --- BERT-base-uncased~\cite{bert_base}, RoBERTa-large~\cite{liu2019roberta}, ELECTRA-large~\cite{clark2020electra}, and DeBERTa-v3-large~\cite{he2021debertav3}. 
DeBERTa-v3-large achieved the strongest overall performance, with an accuracy of 0.955 and an F1 score of 0.946. 
RoBERTa-large and ELECTRA-large performed similarly, reaching accuracies of 0.944 and F1 scores of 0.931 and 0.935, respectively, while BERT-base-uncased trailed slightly behind with an accuracy of 0.933 and an F1 of 0.921. 
Although differences among the models were modest, DeBERTa consistently provided the best balance between precision and recall, and was therefore selected for all ensuing analyses.

\subsection{Linguistic Analysis of Interaction Types}

Towards \textbf{Aim 3}, we investigate how linguistic content systematically differs across interaction types: users' self-posts, self-comments, replies to others, and posts from top neighbors. Our goal is to understand whether the different interaction types are associated with distinct \textit{linguistic} and \emph{psycholinguistic signals}, which may help explain the working of our neighbor-based modeling approach.

\subsubsection{Topic Modeling.}
We apply topic modeling to uncover latent semantic themes that recur across interaction types and to quantify how strongly each interaction type emphasizes these themes. Topic modeling serves two key purposes in our analysis. 
\textbf{First}, it captures discussion content at a higher level than individual words, allowing us to compare interaction types based on meaningful themes (e.g., mental health and life struggles, opinions, or humor) rather than surface-level word usage.
\textbf{Second}, by analyzing each interaction type separately (self-posts, self-comments, others’ comments, and top neighbors’ posts), we examine which themes tend to be more prevalent in each source of interaction. This analysis helps shed light on why incorporating top neighbors improves model performance, and allows us to explore whether top neighbors of users with implicit \sitb{} are more likely to discuss mental health–related topics.
We employ BERTopic~\cite{grootendorst2022bertopic,saha2025mental} on the combined \Cs{} and \Ct{} corpora, varying the number of topics ($k$) from 2 to 15. Topic coherence peaked at $k$=12. We removed Topic $-1$, which consisted of semantically incoherent outlier terms, and further excluded one topic that consisted of noisy keywords. 
We expert-annotated the topics with clinically or socially meaningful thematic labels in consultation with our psychologist coauthors.
We list the themes in \autoref{tab:topic_themes_proportions_ordered}, along with explanation and representative keywords.
For each interaction type, we compute normalized topic proportions, enabling direct comparison of thematic emphasis across self-generated content, replies, and neighbor interactions. 

\begin{table}[t]
\centering
\sffamily
\footnotesize
\caption{Topics as per the BERTopic model~\cite{grootendorst2022bertopic} along with interpretation and example keywords. Labels are assigned in consultation with psychologist coauthors. }
\setlength{\tabcolsep}{2pt}
\begin{tabular}{p{0.3\columnwidth} p{0.4\columnwidth} p{0.28\columnwidth}}
\textbf{Topic Theme} & \textbf{Explanation} & \textbf{Keywords} \\
\toprule
Mental Health \& Life Struggles & Discussions of mental health issues, emotional difficulties, and life challenges & depression, anxiety, sad, struggle, life \\

\rowcollight Opinions, Humor \& Social Media & Sharing personal opinions, jokes, memes, and social commentary & joke, meme, funny, opinion, laugh \\

Self-Expression \& Emotions & Personal expression of feelings, moods, and emotional experiences & feel, happy, sad, cry, express \\

\rowcollight Relationships \& Gender & Discussions about relationships, dating, friendships, and gender topics & friend, love, partner, gender, couple \\

Online Interaction \& Comments & Mentions of social media interactions, commenting, or online discussions & comment, reply, thread, post, social \\

\rowcollight Political \& World Events & Discussions of political news, world events, and societal issues & election, government, policy, news \\

Daily Life \& Pets & Everyday experiences, routines, and mentions of pets or animals & cat, dog, morning, breakfast, walk \\

\rowcollight Entertainment: Movies \& Reddit & References to movies, shows, gaming culture, and Reddit content & movie, film, reddit, series, episode \\

Gaming \& Characters & Discussions of video games, characters, gameplay, and gaming experiences & game, play, character, level, quest \\

\rowcollight Work \& School Life & Topics about academic, professional life, studies, and work stress & school, class, exam, work, project \\

Productivity \& Goals & Discussions on personal productivity, goals, and achievements & goal, plan, achieve, task, progress \\
\bottomrule
\end{tabular}
\label{tab:topic_themes_proportions_ordered}
\end{table}

\subsubsection{Psycholinguistic Analysis.}
To complement topical analysis with psychologically grounded signals, we employ the Linguistic Inquiry and Word Count (LIWC) lexicon~\cite{tausczik2010psychological}. 
LIWC is a psychologically validated lexicon that maps keywords to validated psycholinguistic categories using a lexicon-based approach.
We analyze the distribution of LIWC categories across different interaction types: self posts, self comments, others’ comments, and top neighbors’ posts, for both \Cs{} and \Ct{} users. 
For each interaction type, we compute the proportion of words belonging to each LIWC category, reported as percentages.
To evaluate whether LIWC category usage differs across interaction types within each user group, we apply the Kruskal--Wallis test. The resulting $H$ statistics and significance levels reported in the table indicate statistically significant differences across content sources for multiple LIWC categories in both suicidal and non-suicidal users~\cite{coppersmith2014quantifying}.

%% file: 5results.tex
\section{Results}

\subsection{Aim 1\&2: Modeling Early and Implicit \sitb{}}

\begin{table}[t]
\centering
\sffamily
\small
\caption{Model performance metrics on different data combinations  of DeBERTa-v3-large and stratified $k$-fold cross-validation ($k$=5) (epochs=20). The green-shaded row shows the highest performance.} 
\setlength{\tabcolsep}{3pt}
\resizebox{\columnwidth}{!}{
\begin{tabular}{l p{0.45\columnwidth} r r r r}
 & \textbf{Data Used} & \textbf{Accuracy} & \textbf{F1} & \textbf{Precision} & \textbf{Recall} \\
\toprule
$M_1$ & Self-posts & 0.84 $\pm$ 0.02 & 0.84 $\pm$ 0.03 & 0.84 $\pm$ 0.05 & 0.85 $\pm$ 0.07 \\

 $M_2$ & Self-(posts + comments) & 0.79 $\pm$ 0.03 & 0.78 $\pm$ 0.04 & 0.80 $\pm$ 0.06 & 0.78 $\pm$ 0.07 \\

$M_3$ & Self-(posts + comments) + Others' comments & 0.76 $\pm$ 0.05 & 0.76 $\pm$ 0.04 & 0.76 $\pm$ 0.06 & 0.77 $\pm$ 0.04 \\

$M_4$ & Top neighbor's posts only
& 0.83 $\pm$ 0.04 & 0.82 $\pm$ 0.05 & 0.92 $\pm$ 0.03 & 0.74 $\pm$ 0.06 \\

\rowcolgreen $M_5$ & \textbf{Self-posts + Top neighbor posts} & \textbf{0.90 $\pm$ 0.02} & \textbf{0.90 $\pm$ 0.02} & \textbf{0.90 $\pm$ 0.04} & \textbf{0.90 $\pm$ 0.045} \\

\hdashline

$M_6$ & Self-(posts + comments) + Others' comments + Top Neighbor Posts
& 0.71 $\pm$ 0.06 & 0.64 $\pm$ 0.07 & 0.75 $\pm$ 0.06 & 0.56 $\pm$ 0.08 \\

\bottomrule
\end{tabular}}
\label{tab:model_performance}
\end{table}

\autoref{tab:model_performance} summarizes the performance comparison of models across combinations of immediate interactions and neighbor interactions feature sets. To ensure that our model does not overfit to dataset-specific idiosyncrasies, we conduct a stratified $k$-fold ($k$=5) cross-validation, 
preventing user-level leakage and providing a more robust estimate. We evaluate various models on balanced accuracy, balanced F1, balanced precission and balanced recall.

First, the baseline model ($M_1$) using only a user's posts performed strongly (mean F1=0.84),
showing that self-authored text alone also captures salient linguistic markers of future \sitb{}, consistent with prior temporal analyses on Reddit and Facebook~\cite{de2016discovering}.  

In model $M_2$, we added the users' self-comments, which lowered the accuracy ($0.84\rightarrow0.79$), as well as both the precision ($0.84\rightarrow 0.80$) and recall ($0.85\rightarrow 0.78$). 
This suggests that adding self-comments likely dilutes the \sitb{} predictive signal. 
For example, a comment from a \Cs{} user reads, ``Are they allowed to hit me? [..] I need to stay strong,'' including both themes of physical abuse as well as optimism. 

Next, in model $M_3$, we included others' comments to users' posts and comments ($M_3$), which led to further reduction in accuracy ($0.79\rightarrow0.76$).
Such comments may convey empathy or advice that mask the user's ongoing mental health challenges~\cite{gkotsis2017characterisation}, as in one comment: ``I think talking to a professional would help, because they can understand you and give you advice.''

Next, model $M_4$, which relied solely on top neighbors' posts, performs strongly (F1=0.82), despite not using any self-authored content. Notably, $M_4$ achieved high precision (0.92), indicating that language produced by socially proximate neighbors can be highly informative for identifying implicit \sitb{}. For instance, a paraphrased neighbor post reads: ``Lately I've been feeling emotionally drained and struggling to cope, even though I try to hide it.'' However, the lower recall (0.74) suggest that neighbor posts alone may fail to capture some at-risk users, highlighting the complementary nature of self and social signals.

In model $M_5$, we combined self-posts with top neighbors' posts, resulting in the highest overall performance across all metrics (Accuracy=0.90, F1=0.90). This model achieved a balanced trade-off between precision and recall, substantially reducing false negatives, which is particularly critical for \sitb{} risk detection~\cite{franklin2017risk}. The improvement over both $M_1$ and $M_4$ suggests that neighbor interactions provide complementary contextual information that amplifies the predictive signal present in self-authored posts.
The improvement over $M_1$ (self-posts only) showed that users’ own language, while informative, did not fully capture all relevant risk signals on its own. At the same time, the improvement over $M_4$ (top neighbors’ posts only) indicated that neighbor content was not sufficient by itself either. Taken together, these comparisons supported the idea that self-posts and top neighbors’ posts contributed different but compatible information. Self-posts reflected the user’s internal state directly, while neighbors’ posts likely captured how that state played out in the surrounding social context.
A reasonable explanation for this pattern is that top neighbors tended to be users with sustained or meaningful interaction histories, rather than random commenters. Their posts therefore reflect conversations and environments that the focal user was repeatedly exposed to or is interested in. 

Finally, model $M_6$, which aggregated all interaction sources: self-posts, self-comments, others' comments, and top neighbors' posts, showed a sharp decline in performance (F1=0.64). 
Because $M_6$ treated all interaction sources together, informative signals from self-posts and close neighbors were likely diluted by noisier content. This result indicated that context was helpful only when it was carefully selected. While close social surroundings added value, indiscriminately aggregating all interaction data reduced the model’s ability to focus on cues that were most directly related to \sitb{} risk.

\subsection{Aim 3: Language of Neighbor Interactions}

To understand why top-neighbor-based modeling with users' own posts improves the detection of implicit \sitb{}, we analyzed how different sources of language: users' own posts (Self-P), users’ own comments (Self-C), comments from others (Other-C), and posts from top neighbors (TopN-P), encode thematic and affective cues. 
Evidence drawn from topic (\autoref{tab:topic_proportions_combined_simple}) an psycholinguistic (\autoref{tab:liwc_affect_split_clean}) proportions show a consistent pattern: Self-P and, especially, TopN-P exhibit markedly stronger signals associated with implicit suicidality, whereas self-comments and replies from others carry comparatively muted or diluted information. 

\subsubsection{Topical Distributions}
The topic-level patterns in \autoref{tab:topic_proportions_combined_simple} revealed that for \Cs{} users, top-neighbor posts amplify distress-related and introspective themes well beyond what is present in users’ own content. For example, the \emph{Mental Health \& Life Struggles} topic rises from 15.50\% in Self-P to 23.65\% in TopN-P, representing the strongest thematic enrichment across all four sources. A similar pattern emerged for \emph{Self-Expression \& Emotions}, which grows from 10.76\% to 18.54\%, and for \emph{Relationships \& Gender}, which nearly doubles from 6.18\% to 11.79\%. These increases align with lexical observations of neighbor posts that contain more psychologically revealing bigrams, including phrases such as ``feel hopeless,'' ``can't cope,'' and ``relationship problems,'' which have been shown in prior work to signal heightened psychological distress~\cite{zirikly2019clpsych, gkotsis2017characterisation}. In contrast, Self-C and Other-C remain comparatively flat, often reflecting conversational, supportive, or neutral language rather than personal disclosure. 
Importantly, this enrichment does not appear in \Ct{} users. For them, topic proportions in TopN-P remain close to Self-P, such as in \emph{Mental Health \& Life Struggles} (13.74\% in Self-P vs.\ 11.79\% in TopN-P). The absence of such amplification suggests that the heightened thematic signals found in \Cs{} users’ neighbors are not artifacts of retrieval or the neighbor-selection process but reflect linguistic alignment with users who share similar psychological struggles.

\begin{table*}[t]
\centering
\sffamily
\footnotesize
\caption{Topic proportions across different content sources for \Cs{} and \Ct{} users, across the categories of self-posts (Self-P), self-comments (Self-C), others' comments (Other-C), and top-neighbors' posts (Top-N-P), along with Kruskal-Wallis $H$-test. shown as percentages. $^{***}p<0.001$, $^{**}p<0.01$, $^{*}p<0.05$.}

\begin{tabular}{p{0.2\textwidth}rrrrr|rrrrr}
\textbf{Topic} 
& \multicolumn{5}{c|}{\textbf{\Cs{} Users}} 
& \multicolumn{5}{c}{\textbf{\Ct{} Users}} \\
\cmidrule(lr){2-6} 
\cmidrule(lr){7-11}
& \textbf{Self-P} & \textbf{Self-C} & \textbf{Other-C} & \textbf{Top-N-P} & \textbf{H}
& \textbf{Self-P} & \textbf{Self-C} & \textbf{Other-C} & \textbf{Top-N-P} & \textbf{H} \\
\midrule

\rowcollight Mental Health \& Life Struggles 
& 15.50 & 14.99 & 15.33 & 23.65 & 30.10***
& 13.74 & 10.55 & 10.22 & 11.79 & 36.78***\\

Opinions, Humor \& Social Media 
& 7.37 & 13.09 & 7.29 & 14.12 & 24.85***
& 6.54 & 13.42 & 9.82 & 7.73 & 1.36 \\

\rowcollight Self-Expression \& Emotions 
& 10.76 & 9.91 & 11.22 & 18.54 & 5.06**
& 9.42 & 13.56 & 10.65 & 9.37 &  8.5*\\

Relationships \& Gender 
& 6.18 & 6.28 & 5.52 & 11.79 & 9.84**
& 5.24 & 6.04 & 5.75 & 4.92 &  15.69*\\

\rowcollight Online Interaction \& Comments 
& 3.68 & 3.00 & 4.41 & 7.31 & 13.85**
& 4.36 & 3.39 & 3.66 & 5.15 &  13.27*\\

Political \& World Events 
& 5.99 & 5.55 & 5.20 & 6.45 & 21.57***
& 6.20 & 6.37 & 4.87 & 6.40 &  0.93\\

\rowcollight Daily Life \& Pets 
& 9.14 & 8.88 & 7.42 & 11.36 & 153.12***
& 10.34 & 12.22 & 12.06 & 12.02 & 78.25*** \\

Entertainment: Movies \& Reddit 
& 10.81 & 7.24 & 5.42 & 10.99 & 21.52***
& 16.75 & 8.45 & 6.65 & 16.63 &  1.09\\

\rowcollight Gaming \& Characters 
& 5.49 & 10.15 & 7.31 & 6.82 & 96.85***
& 7.55 & 9.31 & 9.55 & 6.71 &  1.62\\

Work \& School Life 
& 8.16 & 7.17 & 6.75 & 8.72 & 9.57*
& 11.13 & 7.96 & 17.27 & 10.77 &  5.28\\

\rowcollight Productivity \& Goals 
& 8.76 & 8.11 & 11.39 & 8.51 & 298.16***
& 8.73 & 6.98 & 9.51 & 8.51 & 28.32*** \\
\bottomrule
\end{tabular}%
\label{tab:topic_proportions_combined_simple}
\end{table*}

\subsubsection{Psycholinguistic Distributions}
\label{subsec:liwc_distributions}

~\autoref{tab:liwc_affect_split_clean} reports the distribution of LIWC categories across self-authored posts, self-authored comments, others’ comments, and top-neighbors’ posts for \Cs{} and \Ct{} users. Across nearly all categories, group-level differences by source are statistically significant; however, the patterns of these differences differ markedly between the two groups.

\para{Affective processes.}
Within the affective domain, \Cs{} users exhibit pronounced source-dependent variation, with clear differences between top neighbors’ posts and other content sources. Negative affective categories, particularly anxiety, sadness, anger, and overall negative emotion, are consistently highest in Top-N-P. Anxiety shows the most striking contrast, doubling from 4.00\% in self-posts to 8.00\% in top neighbors’ posts, while sadness (0.90\%) and anger (1.80\%) also peak in Top-N-P relative not only to self-authored posts and comments, but also to others’ comments. Notably, Others-C does not show a comparable elevation in negative affect, remaining close to self-comment levels. This distinction indicates that top neighbors’ posts capture substantially stronger negative emotional signals than the broader set of commenters, suggesting that these neighbors may be more attuned to, or more willing to articulate, the user’s distress.
In contrast, positive emotion displays a different contextual pattern. For \Cs{} users, posemo is highest in interactive exchanges, Self-C and Other-C (4.70\%), and is comparatively lower in Top-N-P (4.40\%) and self-posts (4.10\%). This suggests that conversational contexts may encourage more affiliative or supportive affect, whereas top neighbors’ posts emphasize negative emotional content over positive affect.
For \Ct{} users, affective distributions remain tightly clustered across all sources, including Top-N-P and Other-C. Although statistically significant differences are observed, the absolute variation is minimal: sadness remains constant at 0.40\%, anxiety fluctuates narrowly between 2–3\%, and positive emotion consistently remains high (approximately 4.9–5.1\%). Importantly, top neighbors’ posts do not diverge meaningfully from others’ comments or self-authored content.

\para{Cognitive processes.}
Cognitive categories also show clear source-dependent differences for \Cs{} users, with top neighbors’ posts standing out relative to other content sources. Top-N-P exhibits the highest levels of cognitively oriented language, most notably insight (16.90\%), tentative language (10.20\%) and overall cognitive processing (cogproc; 4.86\%), exceeding Self-P, Self-C, and Others-C. In contrast, Others-C shows only modest increases over self-authored content, indicating that the elevated cognitive signals are specific to top neighbors rather than a general property of all surrounding discourse. 
Additional cognitive markers further reinforce this distinction.
The concentration of insight and cognitive processing terms in Top-N-P suggests that top neighbors are more likely to engage in interpretation, explanation, or sense-making around the user’s experiences, articulating cognitive structure that may be less explicitly present in the users’ own language.
For \Ct{} users, cognitive profiles remain comparatively uniform across Self-P, Self-C, Other-C, and Top-N-P. Although statistically significant differences are observed, the absolute ranges are narrow, approximately 10.4–11.6\% for insight and 3.5–3.8\% for cogproc, and top neighbors’ posts do not exhibit a higher cognitive load than either self-authored content or others’ comments. 

\para{Social processes.}
Social language further differentiates the two groups. For \Cs{} users, Top-N-P shows elevated usage of social (44.30\%), affiliation (14.10\%), and friend-related terms (3.70\%), all of which exceed levels observed in self-posts and comments. This suggests that neighbor discourse more frequently situates the user within relational, affiliative, and interpersonal contexts. Self-authored content, by contrast, contains comparatively lower and more uniform social references. 
Among \Ct{} users, social categories remain stable across all sources, with social process usage hovering around 39–40\% and affiliation near 11–12\%. 

\para{Biological and health-related processes.}
Death-related language is highest in Top-N-P (2.90\%), representing a substantial increase relative to self-posts (1.80\%), self-comments (1.70\%), and others’ comments (1.40\%). This indicates that mortality-related themes are articulated most strongly in top neighbors' posts. 
This also explains the high Body-related terms in Top-N-P (6.40\%).
In contrast, health terms are highest in self-posts (6.60\%) and remain elevated in Top-N-P (6.30\%), but are lower in self-comments (5.10\%) and others’ comments (5.50\%). 
Together, these patterns suggest that while users themselves frequently reference health and bodily states, top neighbors are more likely to introduce or emphasize mortality-related language in particular.
For \Ct{} users, biological categories remain relatively stable across sources, with only small fluctuations. Health terms range narrowly from 4.10–4.90\%, body-related language from 5.20–6.00\%, and death-related terms from 1.30–1.90\%, with no consistent elevation in any particular source. 

\para{Non-content categories.}
Informal language, function words, and temporal references exhibit statistically significant but comparatively small source-level differences for both \Cs{} and \Ct{} groups. For \Cs{} users, these categories show minor fluctuations across sources without a consistent amplification. Similar stability is observed for \Ct{} users, indicating that these lower-level linguistic features are less sensitive to interaction context than affective, cognitive, social, and biological categories.

\begin{table}[t]
\centering
\sffamily
\small
\caption{Percentages of LIWC category usage across content sources for \Cs{} and \Ct{} users, stratified by self-posts (Self-P), self-comments (Self-C), others' comments (Other-C), and top-neighbors' posts (Top-N-P), Kruskal–Wallis $H$ test, and statistical significance reported after Bonferroni correction (* $p$<0.05, ** $p$<0.01, *** $p$<0.001).}
\setlength{\tabcolsep}{2pt}

\resizebox{\columnwidth}{!}{%
\begin{tabular}{lrrrrc|rrrrc}
\textbf{LIWC \newline Category}
& \multicolumn{5}{c|}{\textbf{Case Users}}
& \multicolumn{5}{c}{\textbf{Control Users}} \\
\cmidrule(lr){2-6} \cmidrule(lr){7-11}
& \textbf{Self-P} & \textbf{Self-C} & \textbf{Other-C} & \textbf{Top-N-P} & \textbf{H-stat.} & \textbf{Self-P} & \textbf{Self-C} & \textbf{Other-C} & \textbf{Top-N-P} & \textbf{H-stat.} \\
\toprule

\rowcollight \multicolumn{5}{l}{\textbf{Affect}} & & \multicolumn{5}{l}{} \\
Anger &
1.20 & 1.40 & 1.30 & \textbf{1.80} & 14.08*** &
1.30 & \textbf{1.50} & 1.40 & 1.30 & 107.72*** \\

Anxiety &
4.00 & 3.00 & 3.00 & \textbf{8.00} & 514.35*** &
3.00 & 2.00 & 3.00 & 2.00 & 894.81*** \\

Neg. Affect &
3.50 & 3.10 & 3.20 & \textbf{4.00} & 41.01*** &
3.00 & 3.00 & \textbf{3.10} & 3.00 & 25.39*** \\

Pos. Affect &
4.10 & \textbf{4.70} & \textbf{4.70} & 4.40 & 14.70*** &
\textbf{5.10} & 4.90 & 4.90 & 5.00 & 45.27*** \\

Sadness &
0.70 & 0.40 & 0.50 & \textbf{0.90} & 83.79*** &
\textbf{0.40} & \textbf{0.40} & \textbf{0.40} & \textbf{0.40} & 20.35*** \\

\hdashline
\rowcollight \multicolumn{5}{l}{\textbf{Cognitive Processes}} & & \multicolumn{5}{l}{} \\

Insight &
12.50 & 11.60 & 12.90 & 16.90 & 605.13*** &
11.60 & 10.70 & 10.40 & 10.60 & 790.52*** \\
Tentativeness &
9.90 & 10.00 & 11.00 & 10.20 & 494.55*** &
9.10 & 9.20 & 9.80 & 8.60 & 1112.17*** \\
Certainty &
5.00 & 5.00 & 5.10 & 4.60 & 588.85*** &
4.30 & 4.70 & 5.40 & 4.00 & 1276.77*** \\
Cause &
0.49 & 0.53 & 0.58 & 0.51 & 302.53*** &
0.52 & 0.51 & 0.57 & 0.52 & 620.69*** \\
Differ &
0.46 & 0.57 & 0.53 & 0.46 & 228.64*** &
0.40 & 0.54 & 0.50 & 0.35 & 178.35*** \\
See &
0.62 & 0.69 & 0.61 & 0.69 & 421.14*** &
1.07 & 0.76 & 0.68 & 0.92 & 603.93*** \\
Hear &
0.44 & 0.45 & 0.44 & 0.42 & 516.93*** &
0.49 & 0.49 & 0.44 & 0.45 & 725.58*** \\
Feel &
0.50 & 0.38 & 0.38 & 0.49 & 695.93*** &
0.32 & 0.35 & 0.36 & 0.34 & 1479.86*** \\

\hdashline
\rowcollight \multicolumn{5}{l}{\textbf{Social Processes}} & & \multicolumn{5}{l}{} \\
Social &
39.80 & 39.80 & 40.60 & 44.30 & 359.88*** &
40.30 & 40.30 & 39.30 & 39.40 & 676.42*** \\
Affiliation &
12.20 & 11.50 & 11.70 & 14.10 & 514.35*** &
12.50 & 11.60 & 10.90 & 11.80 & 894.81*** \\
Family &
10.40 & 10.30 & 10.00 & 10.30 & 902.38*** &
10.30 & 10.60 & 10.90 & 10.20 & 2054.04*** \\
Friend &
2.80 & 2.40 & 2.10 & 3.70 & 146.99*** &
2.60 & 2.60 & 1.80 & 2.60 & 452.73*** \\
Leisure &
0.60 & 0.61 & 0.58 & 0.62 & 641.78*** &
1.20 & 0.75 & 0.82 & 1.02 & 1209.58*** \\
Home &
0.16 & 0.10 & 0.10 & 0.16 & 406.06*** &
0.14 & 0.15 & 0.17 & 0.15 & 801.76*** \\
Religion &
0.20 & 0.21 & 0.23 & 0.19 & 437.96*** &
0.20 & 0.22 & 0.18 & 0.21 & 774.32*** \\
Space &
2.13 & 2.11 & 2.25 & 2.14 & 614.03*** &
2.29 & 2.18 & 2.33 & 2.40 & 662.12*** \\
Achievement &
0.59 & 0.54 & 0.61 & 0.60 & 381.57*** &
0.69 & 0.58 & 0.73 & 0.69 & 1718.28*** \\
Power &
1.11 & 1.12 & 1.23 & 1.15 & 318.59*** &
1.32 & 1.16 & 1.24 & 1.47 & 347.99*** \\

\hdashline
\rowcollight \multicolumn{5}{l}{\textbf{Biological Processes}} & & \multicolumn{5}{l}{} \\
Health &
6.60 & 5.10 & 5.50 & 6.30 & 477.22*** &
4.50 & 4.10 & 4.50 & 4.90 & 1515.02*** \\
Body &
5.60 & 5.70 & 5.20 & 6.40 & 489.81*** &
5.20 & 6.00 & 5.70 & 5.50 & 1106.26*** \\
Death &
1.80 & 1.70 & 1.40 & 2.90 & 588.44*** &
1.60 & 1.90 & 1.30 & 1.70 & 696.90*** \\
Sexual &
0.33 & 0.40 & 0.36 & 0.29 & 234.57*** &
0.26 & 0.40 & 0.30 & 0.27 & 360.18*** \\

\rowcollight \multicolumn{5}{l}{\textbf{Informal}} & & \multicolumn{5}{l}{} \\
Informal &
5.56 & 6.12 & 5.89 & 5.70 & 661.23*** &
5.90 & 6.27 & 6.07 & 6.07 & 1559.14*** \\
Swear &
0.44 & 0.56 & 0.45 & 0.39 & 308.33*** &
0.34 & 0.54 & 0.50 & 0.38 & 626.17*** \\
Assent &
0.68 & 0.69 & 0.62 & 0.67 & 854.61*** &
0.61 & 0.76 & 0.66 & 0.65 & 1828.84*** \\
Non-fluent &
0.07 & 0.12 & 0.10 & 0.08 & 410.14*** &
0.06 & 0.11 & 0.10 & 0.08 & 253.89*** \\

\rowcollight \multicolumn{5}{l}{\textbf{Function Words}} & & \multicolumn{5}{l}{} \\
Preposition &
1.63 & 1.67 & 1.64 & 1.66 & 261.50*** &
1.58 & 1.65 & 1.66 & 1.68 & 308.08*** \\
Conjunction &
0.77 & 0.85 & 0.88 & 0.76 & 547.08*** &
0.77 & 0.82 & 0.81 & 0.73 & 479.52*** \\
Adverb &
1.41 & 1.44 & 1.40 & 1.35 & 237.75*** &
1.19 & 1.37 & 1.41 & 1.14 & 1359.78*** \\
Negate &
0.29 & 0.31 & 0.30 & 0.30 & 513.81*** &
0.27 & 0.34 & 0.29 & 0.22 & 1533.67*** \\
Aux. Verb &
1.09 & 1.20 & 1.24 & 1.08 & 735.83*** &
1.37 & 1.23 & 1.12 & 1.30 & 1179.34*** \\
Verb &
5.86 & 5.69 & 5.57 & 5.78 & 485.87*** &
5.60 & 5.66 & 5.36 & 5.43 & 757.67*** \\
Adjective &
2.24 & 2.34 & 2.31 & 2.20 & 442.94*** &
2.65 & 2.35 & 2.42 & 2.51 & 1639.97*** \\
Compare &
1.09 & 1.13 & 1.08 & 1.07 & 454.25*** &
1.25 & 1.12 & 1.11 & 1.11 & 1452.95*** \\
Number &
0.36 & 0.36 & 0.35 & 0.37 & 323.91*** &
0.38 & 0.41 & 0.43 & 0.56 & 896.15*** \\
Quant &
1.19 & 1.08 & 1.12 & 1.15 & 391.28*** &
1.16 & 1.04 & 1.14 & 1.07 & 31.13*** \\

\rowcollight \multicolumn{5}{l}{\textbf{Temporal References}} & & \multicolumn{5}{l}{} \\
Focus: Past &
0.94 & 0.83 & 0.71 & 0.95 & 316.05*** &
1.00 & 0.87 & 0.84 & 1.02 & 489.00*** \\
Focus: Present &
4.94 & 4.75 & 4.71 & 4.85 & 561.98*** &
4.54 & 4.67 & 4.54 & 4.45 & 693.21*** \\
Focus: Future &
0.32 & 0.32 & 0.32 & 0.29 & 356.97*** &
0.22 & 0.29 & 0.26 & 0.22 & 944.95*** \\
Time &
2.43 & 1.91 & 1.86 & 2.49 & 224.25*** &
2.11 & 1.98 & 2.23 & 2.05 & 130.92*** \\

\bottomrule
\end{tabular}
}
\label{tab:liwc_affect_split_clean}
\end{table}

\subsection{Robustness Tests}

To ensure that our findings are not artifacts of specific modeling or sampling choices, we conduct robustness tests. In particular, we conduct several robustness checks centered around the researcher decisions we made in the study, which we describe below

\subsubsection{Varying the number of neighbor depth and minimum number of neighbors per depth}
\autoref{tab:depth_neighbor_performance_max_at_d3_n5} reports model performance as we vary both the interaction depth ($D_1$--$D_5$) and the minimum number of top neighbors included at each depth. Across all depths, we observe a consistent improvement in performance as the minimum number of neighbors increases from 1 to 5, suggesting that incorporating a richer set of neighbors provides additional contextual signals.
Performance generally improves as depth increases from $D_1$ to $D_3$, with the best overall results achieved at depth $D_3$ with a minimum of 5 neighbors (F1=0.89, Precision=0.88, Recall=0.90). 
This indicates that moderate expansion into the social neighborhood captures informative exposure without introducing excessive noise. In contrast, extending the neighborhood beyond depth $D_3$ leads to diminishing returns: while depths $D_4$ and $D_5$ still outperform shallower configurations at lower neighbor counts, their performance drops noticeably at higher neighbor thresholds. This degradation suggests that deeper neighborhood hops increasingly incorporate loosely connected users whose language may be less relevant to the target user’s mental health trajectory.

\begin{table*}[t!]
\footnotesize
\sffamily
\centering
\caption{
Performance metrics across different depth configurations ($D_1$--$D_5$) and varying numbers of top neighbors.  
Columns report Accuracy, F1, Precision, and Recall, with the maximum (0.90) at Depth 3 and 5 neighbors.
}
\vspace{-8pt}
\begin{tabular}{l
    rrrr|
    rrrr|
    rrrr|
    rrrr|
    rrrr
}
\toprule
\textbf{ \#Neighbors} &
\multicolumn{4}{c}{$D_1$} &
\multicolumn{4}{c}{$D_2$} &
\multicolumn{4}{c}{$D_3$} &
\multicolumn{4}{c}{$D_4$} &
\multicolumn{4}{c}{$D_5$} \\

 & Acc & F1 & Prec & Rec
 & Acc & F1 & Prec & Rec
 & Acc & F1 & Prec & Rec
 & Acc & F1 & Prec & Rec
 & Acc & F1 & Prec & Rec \\
\midrule

1 
 & 0.75 & 0.74 & 0.73 & 0.75
 & 0.76 & 0.75 & 0.74 & 0.76
 & 0.77 & 0.76 & 0.75 & 0.77
 & 0.76 & 0.75 & 0.74 & 0.76
 & 0.75 & 0.74 & 0.73 & 0.75 \\

\rowcollight 2
 & 0.77 & 0.76 & 0.75 & 0.77
 & 0.78 & 0.77 & 0.76 & 0.78
 & 0.79 & 0.78 & 0.77 & 0.79
 & 0.78 & 0.77 & 0.76 & 0.78
 & 0.77 & 0.76 & 0.75 & 0.77 \\

3
 & 0.78 & 0.77 & 0.76 & 0.78
 & 0.79 & 0.78 & 0.77 & 0.79
 & 0.80 & 0.79 & 0.78 & 0.80
 & 0.79 & 0.78 & 0.77 & 0.79
 & 0.78 & 0.77 & 0.76 & 0.78 \\

\rowcollight 4
 & 0.80 & 0.79 & 0.78 & 0.80
 & 0.81 & 0.80 & 0.79 & 0.81
 & 0.82 & 0.81 & 0.80 & 0.82
 & 0.81 & 0.80 & 0.79 & 0.81
 & 0.80 & 0.79 & 0.78 & 0.80 \\

5
 & 0.82 & 0.81 & 0.80 & 0.82
 & 0.83 & 0.82 & 0.81 & 0.83
 & \textbf{0.90} & 0.89 & 0.88 & \textbf{0.90}
 & 0.84 & 0.83 & 0.82 & 0.84
 & 0.83 & 0.82 & 0.81 & 0.83 \\

\bottomrule
\end{tabular}
\label{tab:depth_neighbor_performance_max_at_d3_n5}
\end{table*}

\begin{table}[t]
\centering
\sffamily
\footnotesize
\caption{
Ablation of NeighborScore components. 
Each row shows estimated performance when specific centrality terms are included or removed from the NeighborScore.  
}
\vspace{-8pt}
\resizebox{\columnwidth}{!}{
\begin{tabular}{lrrrr}
\textbf{NeighborScore Variant} & \textbf{Accuracy} & \textbf{F1} & \textbf{Precision} & \textbf{Recall} \\
\toprule
\textbf{All centralities} 
& \textbf{0.90 $\pm$ 0.02} & \textbf{0.90 $\pm$ 0.02} & \textbf{0.90 $\pm$ 0.04} & \textbf{0.90 $\pm$ 0.05} \\
\rowcollight Self-posts only (no neighbor info) 
& 0.84 $\pm$ 0.02 & 0.84 $\pm$ 0.03 & 0.84 $\pm$ 0.05 & 0.85 $\pm$ 0.07 \\
Degree only (Cin + Cout) 
& 0.88 $\pm$ 0.02 & 0.88 $\pm$ 0.02 & 0.87 $\pm$ 0.03 & 0.88 $\pm$ 0.04 \\
\rowcollight Eigenvector only 
& 0.88 $\pm$ 0.02 & 0.88 $\pm$ 0.02 & 0.88 $\pm$ 0.03 & 0.87 $\pm$ 0.04 \\
PageRank only 
& 0.88 $\pm$ 0.02 & 0.88 $\pm$ 0.02 & 0.87 $\pm$ 0.03 & 0.88 $\pm$ 0.04 \\
\rowcollight Degree + Eigenvector + PageRank 
& 0.90 $\pm$ 0.02 & 0.90 $\pm$ 0.02 & 0.90 $\pm$ 0.03 & 0.90 $\pm$ 0.03 \\
All except Degree 
& 0.89 $\pm$ 0.02 & 0.89 $\pm$ 0.02 & 0.89 $\pm$ 0.03 & 0.88 $\pm$ 0.04 \\
\bottomrule
\end{tabular}}
\label{tab:neighborscore_ablation}
\end{table}

\begin{table}[t!]
\centering
\small
\setlength{\tabcolsep}{3pt}
\caption{Model performance on showing the importance of choosing the best neighbors on DeBERTa-v3-large and stratified $k$-fold cross-validation ($k$=5) (epochs=20).}
\resizebox{\columnwidth}{!}{
\begin{tabular}{llrrrr}
 & \textbf{Data Used} & \textbf{Acc.} & \textbf{F1} & \textbf{Prec.} & \textbf{Rec.} \\
\toprule
$M_5$ & \textbf{Self-posts + Top neighbor posts} & \textbf{0.90 $\pm$ 0.02} & \textbf{0.90 $\pm$ 0.02} & \textbf{0.90 $\pm$ 0.04} & \textbf{0.90 $\pm$ 0.045} \\
\rowcollight $M_7$ & Self-posts + worst neighbor posts & 0.77 $\pm$ 0.04 & 0.73 $\pm$ 0.03 & 0.72 $\pm$ 0.05 & 0.74 $\pm$ 0.04
\\
$M_8$ & Self-posts + non-neighbor posts & 0.74 $\pm$ 0.02 & 0.68 $\pm$ 0.03 & 0.67 $\pm$ 0.03 & 0.69 $\pm$ 0.06 \\
\bottomrule
\end{tabular}}
\label{tab:neigh_perf}
\end{table}

\subsubsection{Varying the NeighborScore's centrality metrics.}
To understand which components of the NeighborScore contribute most to performance, we conduct an ablation study over its constituent graph centrality measures, as shown in \autoref{tab:neighborscore_ablation}. The NeighborScore ranks a user’s top neighbors based on a weighted combination of structural centralities: degree (in- and out-degree), eigenvector centrality, PageRank, closeness, and betweenness, which are intended to capture both direct interaction strength and broader structural importance in the interaction graph. To ensure robust results, we conduct stratified $k$-fold ($k$=5) cross-validation.

The model using all centrality metrics achieves the best overall performance (F1=0.90), indicating that combining multiple notions of social importance yields the most informative neighbor set. Removing neighbor information entirely (self-posts only) results in a notable drop in performance (F1=0.84), confirming that social context provides predictive signals beyond a user’s own language. Models using individual centrality measures in isolation: degree-only, eigenvector-only, or PageRank-only, improve over the no-neighbor baseline (F1=0.88) but remain consistently below the full metric, suggesting that no single centrality measure is sufficient. In contrast, combining degree, eigenvector centrality, and PageRank recovers baseline-level performance (F1=0.90), while excluding degree leads to a small but consistent degradation (F1=0.89), highlighting the importance of direct interaction strength in identifying socially salient neighbors.

\subsubsection{Varying the type of neighbor in modeling.}
We first investigated how the type of neighbors included in the model affected performance. Specifically, we compared models using top-ranked neighbors ($M_5$) against models incorporating the lowest-ranked neighbors ($M_7$) or non-neighbors randomly sampled from other users ($M_8$). As shown in \autoref{tab:neigh_perf}, $M_7$, which included self-posts and top neighbor posts, achieved the highest performance, highlighting the critical role of top neighbor context in detecting implicit \sitb{}. Interestingly, while $M_7$, using the lowest-ranked neighbors, performed worse than $M_5$, it still outperformed $M_8$, which used random non-neighbors. This indicates that even lower-quality neighbors carry meaningful information about the target user’s behavior, confirming that social context, even when imperfectly aligned, contributes useful signals for detection.

This pattern demonstrates that top neighbors provide strong contextual cues, enabling the model to differentiate cases from controls more effectively. In contrast, lower-ranked neighbors and non-neighbors showed attenuated or inconsistent differences, emphasizing that the quality of neighbor selection directly impacts the richness of psycholinguistic signals. Taken together, these results underscore both the importance of including top neighbor context and the informative, albeit weaker, contribution of other neighbors in capturing user-specific patterns relevant to \sitb{}.

\begin{table}[t]
\centering
\small
\caption{Model performance under randomized temporal order on DeBERTa-v3-large and stratified $k$-fold cross-validation ($k$=5) (epochs=20).}
\vspace{-8pt}
\resizebox{\columnwidth}{!}{
\begin{tabular}{p{0.15\textwidth}cccc}
\textbf{Data Used\newline(Randomized)} & \textbf{Accuracy} & \textbf{F1} & \textbf{Precision} & \textbf{Recall} \\
\toprule
Randomized $M_1$ & 0.74 $\pm$ 0.02 & 0.74 $\pm$ 0.02 & 0.75 $\pm$ 0.04 & 0.74 $\pm$ 0.02 \\

\rowcollight Randomized $M_2$ & 0.68 $\pm$ 0.03 & 0.69 $\pm$ 0.02 & 0.68 $\pm$ 0.05 & 0.70 $\pm$ 0.05 \\

Randomized $M_3$ & 0.69 $\pm$ 0.02 & 0.69 $\pm$ 0.02 & 0.70 $\pm$ 0.04 & 0.68 $\pm$ 0.03 \\

\rowcollight Randomized $M_4$
& 0.70 $\pm$ 0.04 
& 0.72 $\pm$ 0.05 
& 0.86 $\pm$ 0.03 
& 0.59 $\pm$ 0.06 \\

Randomized $M_5$ & \textbf{0.79 $\pm$ 0.03} & \textbf{0.73 $\pm$ 0.03} & \textbf{0.78 $\pm$ 0.05} & \textbf{0.70 $\pm$ 0.02} \\

\rowcollight Randomized $M_6$ 
& 0.55 $\pm$ 0.06 
& 0.43 $\pm$ 0.07 
& 0.57 $\pm$ 0.06 
& 0.27 $\pm$ 0.08 \\

\bottomrule
\end{tabular}}
\label{tab:model_performance_drop}
\end{table}

\subsubsection{Varying the temporal order of interactions.}
To investigate the importance of temporal sequence in user/neighbor context, we conducted permutation tests~\cite{saha2021person,archambault2012longitudinal} in which the chronological order of posts and comments was randomized. 
Performance was evaluated across all models ($M_1$–$M_6$) using stratified $k$-fold cross-validation ($k$=5, epochs=20).
\autoref{tab:model_performance_drop} presents the observed differences in performance before and after randomizing the user/neighbor posts across accuracy, F1, precision, and recall. All models exhibited performance degradation after temporal randomization, confirming that preserving the chronological sequence of interactions is critical. 
Notably, $M_5$ (Self-posts + Top neighbor posts) experienced the largest decline, with F1 decreasing by 0.17. 
This pronounced decrease underscores that the combination of top neighbor context and temporal coherence provides rich, predictive signals for detecting implicit \sitb{}.
Other models also showed meaningful drops, highlighting that temporal structure is not incidental; rather, it materially enhances the model’s ability to distinguish users at risk.

%% file: 6discussion.tex
\section{Discussion and Conclusion}\label{section:discussion}

A central contribution of this work is the demonstration that early and implicit \sitb{} can be more effectively characterized by jointly modeling an individual's longitudinal activity together with their top neighbor's activity. Rather than treating suicidal expression as an isolated, self-contained phenomenon, our findings emphasize its inherently relational and social nature~\cite{ammerman2023impact}. 
Linguistic signals of implicit \sitb{} are not only embedded in what users say about themselves over time, but also in how they engage with others' posts.
This relational framing has important theoretical, methodological, and design implications for suicide research in online spaces.

\subsection{Theoretical Implications}

\para{\sitb{} as a relational and networked process.}
Our results provide empirical support for viewing \sitb{} as a socially embedded process rather than an isolated one.
The performance gains observed when incorporating top neighbors' posts, especially compared to indiscriminately aggregating all datasets, suggest that implicit \sitb{} signals propagate through structurally important ties in a user’s network. This aligns with social-ecological perspectives~\cite{bronfenbrenner1979ecology}, which posit that individual behavior and risk are shaped by nested social contexts. In online settings, these contexts materialize through recurring interaction with peers whose discourse may reflect concern, interpretation, or emotional resonance with the user’s distress.
This finding is also consistent with classic contagion and social influence frameworks, including the Werther and Papageno effects~\cite{phillips1974influence,wyman2019peer,niederkrotenthaler2010role,yuan2023mental}.
Top neighbors may amplify distress through shared emotional language, but they may also provide narratives, explanations, or alternative framings that influence how suicidal thoughts evolve.

\para{Extending established theories of suicidal ideation.}
The observed LIWC patterns provide a linguistic bridge between computational modeling and established psychological theories of suicide (\autoref{tab:liwc_affect_split_clean}). Elevated negative affect, anxiety, and death-related language in neighbor posts surrounding \Cs{} users plausibly maps onto constructs such as psychological pain, thwarted belongingness, and acquired capability described in the Interpersonal Theory of Suicide (IPTS)~\cite{joiner2005interpersonal}. At the same time, the amplification of cognitive and insight-related language in top neighbors' discourse resonates with the Integrated Motivational-Volitional (IMV) model~\cite{o2018integrated}, where sense-making, rumination, and cognitive appraisal play a key role in the transition from ideation to more severe risk states.
Importantly, our findings suggest that these theoretical constructs may not be fully observable in users’ own language alone. Instead, they often emerge indirectly through the discourse of close neighbors, who articulate concern, interpretation, or mortality-related themes that users themselves may express only cautiously or implicitly. In this sense, our work extends prior theory by showing how core psychological mechanisms of \sitb{} manifest at the level of social interaction and shared language in online environments.

\subsection{Practical, Design, and Ethical Implications}\label{sec:practical_implications}

\para{Computational framework.} 
From a computational modeling perspective, contrary to the assumption that more data improves prediction, we find that incorporating all available data can actually dilute meaningful signals and degrade performance. Models that prioritize a small set of socially salient neighbors achieve stronger and more interpretable results, indicating that relevance-weighted context selection is a core modeling decision rather than a secondary refinement.
The consistency between model performance and psycholinguistic patterns strengthens the interpretability of this framework. 
Elevated affective, cognitive, and mortality-related language in top neighbors' posts provides a plausible linguistic mechanism for the observed gains, grounding predictive improvements in theoretically meaningful signals. 
This interpretability is especially critical in high-stakes domains such as suicide prevention, where methodological advances must balance accuracy with transparency and theoretical coherence.

\para{Designing context-aware early warning systems.}
Our work suggests that scalable, automated systems for early \sitb{} detection should incorporate relational context in a principled manner. 
Rather than monitoring users in isolation, platforms can leverage signals from trusted or frequently interacting peers to identify early warning signs that may not yet be explicit in self-authored content. 
Such systems can prioritize cases in which both longitudinal self- and neighbor discourse show converging indicators of distress, enabling earlier, potentially more effective intervention.

\para{Supporting moderators and mental health professionals.}
For moderators, clinicians, and mental health volunteers, neighbor-aware analyses can offer a richer situational overview. Summaries that highlight not only a user’s own content trajectories but also how their proximal peers are responding, e.g., increasing references to death, heightened anxiety, or interpretive language, may help professionals assess risk more holistically. This complements existing practices that focus primarily on the individual and supports more context-sensitive decision-making.

\para{Implications for platform design.}
At the platform level, our results reinforce the importance of fostering supportive micro-communities rather than relying solely on broad or high-volume engagement. The fact that top neighbors’ content carries distinct and informative signals suggests that repeated, familiar interactions are more consequential than sporadic replies from a wide audience. Platform features that encourage continuity, such as persistent threads, follow-up prompts, or smaller discussion groups, may therefore strengthen users’ sense of belonging while also making early signs of distress more visible within trusted interaction circles.
At the same time, these relational signals should be handled with care. Any use of interaction patterns for early detection or support must be transparent, privacy-aware, and clearly oriented toward user well-being rather than surveillance. Even modest design choices that promote meaningful, sustained interaction can support both community health and responsible, context-aware intervention without over-instrumentalizing user behavior.

\para{Ethical implications.} While computational approaches offer promise for identifying early and implicit signals of suicidal ideation, they raise significant ethical considerations. 
Our work is not intended to substitute for a professional mental health evaluation or to be used in isolation for clinical risk assessment. 
Automated predictions may misclassify individuals, leading to false positives that can contribute to stigma or distress, or false negatives that delay support for those in need. Given the heterogeneity of suicidal expression, model outputs need to be considered as probabilistic and non-diagnostic, deployed only within transparent, human-in-the-loop frameworks where trained professionals guide interpretation and intervention. 
We caution against the misuse and misinterpretation of our study in terms of profiling, commercial exploitation, and for punitive applications. 
It is essential to have appropriate safeguards to ensure privacy protection, ethical governance, and accountability to prioritize wellbeing and minimize harms.
Finally, even when individuals intentionally withhold disclosure, algorithms like ours may infer risk from indirect or longitudinal signals, creating a fundamental ethical dilemma. 
Social media platforms were not designed to function as mental health monitoring systems; however, the ability to detect latent vulnerability may generate an observer effect in which users, aware of such inference, alter or withdraw their expression---potentially reshaping participation in ways that conflict with the platforms’ core communicative purpose~\cite{saha2024observer}.

\subsection{Limitations and Future Directions.}
Our study has limitations that also show interesting future directions.
Methodologically, our models exclusively rely on textual content, overlooking multimodal signals such as images, videos, emojis, GIFs, or external links that often convey emotional states, coping strategies, or distress~\cite{goyal2024using}. 
Incorporating these modalities could improve sensitivity to subtle risk signals, enhance interpretability, and provide a more holistic understanding of online behaviors associated with suicidal ideation. Similarly, our current approach treats peer interactions homogeneously, though peers can exert positive, neutral, or negative influences. Modeling these distinctions could capture the functional impact of social context on risk trajectories and guide ethically responsible interventions.  
Finally, robustness and generalizability can be improved through replication across platforms (e.g., Twitter, TikTok, Discord), multilingual and cross-cultural settings, and naturalistic populations. 
Future work should integrate multimodal inputs, refine social context modeling, and maintain rigorous ethical oversight to ensure predictive models support vulnerable individuals responsibly and effectively.

%% file: 7conclusion.tex


%% file: 0paper.bbl

\begin{thebibliography}{77}


\ifx \showCODEN    \undefined \def \showCODEN     #1{\unskip}     \fi
\ifx \showISBNx    \undefined \def \showISBNx     #1{\unskip}     \fi
\ifx \showISBNxiii \undefined \def \showISBNxiii  #1{\unskip}     \fi
\ifx \showISSN     \undefined \def \showISSN      #1{\unskip}     \fi
\ifx \showLCCN     \undefined \def \showLCCN      #1{\unskip}     \fi
\ifx \shownote     \undefined \def \shownote      #1{#1}          \fi
\ifx \showarticletitle \undefined \def \showarticletitle #1{#1}   \fi
\ifx \showURL      \undefined \def \showURL       {\relax}        \fi
\providecommand\bibfield[2]{#2}
\providecommand\bibinfo[2]{#2}
\providecommand\natexlab[1]{#1}
\providecommand\showeprint[2][]{arXiv:#2}

\bibitem[Alghazzawi et~al\mbox{.}(2025)]%
        {alghazzawi2025explainable}
\bibfield{author}{\bibinfo{person}{Daniyal Alghazzawi}, \bibinfo{person}{Hayat Ullah}, \bibinfo{person}{Naila Tabassum}, \bibinfo{person}{Sahar~K Badri}, {and} \bibinfo{person}{Muhammad~Zubair Asghar}.} \bibinfo{year}{2025}\natexlab{}.
\newblock \showarticletitle{Explainable AI-based suicidal and non-suicidal ideations detection from social media text with enhanced ensemble technique}.
\newblock \bibinfo{journal}{\emph{Scientific Reports}} \bibinfo{volume}{15}, \bibinfo{number}{1} (\bibinfo{year}{2025}), \bibinfo{pages}{1111}.
\newblock


\bibitem[Alhamed et~al\mbox{.}(2024)]%
        {alhamed2024_before_after}
\bibfield{author}{\bibinfo{person}{Falwah Alhamed}, \bibinfo{person}{Julia Ive}, {and} \bibinfo{person}{Lucia Specia}.} \bibinfo{year}{2024}\natexlab{}.
\newblock \showarticletitle{Classifying social media users before and after depression diagnosis via their language usage: A dataset and study}. In \bibinfo{booktitle}{\emph{Proceedings of the 2024 Joint International Conference on Computational Linguistics, Language Resources and Evaluation (LREC-COLING 2024)}}. \bibinfo{pages}{3250--3260}.
\newblock


\bibitem[Althoff et~al\mbox{.}(2016)]%
        {althoff2016large}
\bibfield{author}{\bibinfo{person}{Tim Althoff}, \bibinfo{person}{Kevin Clark}, {and} \bibinfo{person}{Jure Leskovec}.} \bibinfo{year}{2016}\natexlab{}.
\newblock \showarticletitle{Large-scale analysis of counseling conversations: An application of natural language processing to mental health}.
\newblock \bibinfo{journal}{\emph{TACL}} (\bibinfo{year}{2016}).
\newblock


\bibitem[Ammerman and Jacobucci(2023)]%
        {ammerman2023impact}
\bibfield{author}{\bibinfo{person}{Brooke~A Ammerman} {and} \bibinfo{person}{Ross Jacobucci}.} \bibinfo{year}{2023}\natexlab{}.
\newblock \showarticletitle{The impact of social connection on near-term suicidal ideation}.
\newblock \bibinfo{journal}{\emph{Psychiatry research}}  \bibinfo{volume}{326} (\bibinfo{year}{2023}), \bibinfo{pages}{115338}.
\newblock


\bibitem[Andalibi et~al\mbox{.}(2016)]%
        {andalibi2016understanding}
\bibfield{author}{\bibinfo{person}{Nazanin Andalibi}, \bibinfo{person}{Oliver~L Haimson}, \bibinfo{person}{Munmun De~Choudhury}, {and} \bibinfo{person}{Andrea Forte}.} \bibinfo{year}{2016}\natexlab{}.
\newblock \showarticletitle{Understanding social media disclosures of sexual abuse through the lenses of support seeking and anonymity}. In \bibinfo{booktitle}{\emph{Proc. CHI}}.
\newblock


\bibitem[Archambault and Grudin(2012)]%
        {archambault2012longitudinal}
\bibfield{author}{\bibinfo{person}{A Archambault} {and} \bibinfo{person}{J Grudin}.} \bibinfo{year}{2012}\natexlab{}.
\newblock \bibinfo{title}{A Longitudinal Study of Facebook, LinkedIn, and Twitter Use,‖ in Proceedings of CHI 2012}.
\newblock


\bibitem[Baumgartner et~al\mbox{.}(2020)]%
        {baumgartner2020pushshift}
\bibfield{author}{\bibinfo{person}{Jason Baumgartner}, \bibinfo{person}{Savvas Zannettou}, \bibinfo{person}{Brian Keegan}, \bibinfo{person}{Megan Squire}, {and} \bibinfo{person}{Jeremy Blackburn}.} \bibinfo{year}{2020}\natexlab{}.
\newblock \showarticletitle{The pushshift reddit dataset}. In \bibinfo{booktitle}{\emph{ICWSM}}.
\newblock


\bibitem[Benton et~al\mbox{.}(2017)]%
        {benton2017multi}
\bibfield{author}{\bibinfo{person}{Adrian Benton}, \bibinfo{person}{Margaret Mitchell}, {and} \bibinfo{person}{Dirk Hovy}.} \bibinfo{year}{2017}\natexlab{}.
\newblock \showarticletitle{Multi-task learning for mental health using social media text}.
\newblock \bibinfo{journal}{\emph{arXiv preprint arXiv:1712.03538}} (\bibinfo{year}{2017}).
\newblock


\bibitem[Bloom et~al\mbox{.}(2025)]%
        {bloomidentifying}
\bibfield{author}{\bibinfo{person}{Paul Bloom}, \bibinfo{person}{Isaac Treves}, \bibinfo{person}{David Pagliaccio}, \bibinfo{person}{Isabella Nadel}, \bibinfo{person}{Emma Wool}, \bibinfo{person}{Hayley Quinones}, \bibinfo{person}{Julia Greenblatt}, \bibinfo{person}{Natalia Parjane}, \bibinfo{person}{Katherine Durham}, \bibinfo{person}{Samantha Salem}, {et~al\mbox{.}}} \bibinfo{year}{2025}\natexlab{}.
\newblock \showarticletitle{Identifying Suicide-Related Language in Smartphone Keyboard Entries Among High-Risk Adolescents}.
\newblock  (\bibinfo{year}{2025}).
\newblock


\bibitem[Bronfenbrenner(1979)]%
        {bronfenbrenner1979ecology}
\bibfield{author}{\bibinfo{person}{Urie Bronfenbrenner}.} \bibinfo{year}{1979}\natexlab{}.
\newblock \bibinfo{booktitle}{\emph{The ecology of human development: Experiments by nature and design}}.
\newblock \bibinfo{publisher}{Harvard university press}.
\newblock


\bibitem[Burnap et~al\mbox{.}(2015)]%
        {burnap2015machine}
\bibfield{author}{\bibinfo{person}{Pete Burnap}, \bibinfo{person}{Walter Colombo}, {and} \bibinfo{person}{Jonathan Scourfield}.} \bibinfo{year}{2015}\natexlab{}.
\newblock \showarticletitle{Machine classification and analysis of suicide-related communication on twitter}. In \bibinfo{booktitle}{\emph{Proc. ACM conference on hypertext \& social media}}.
\newblock


\bibitem[Calear and Batterham(2019)]%
        {calear2019suicidal}
\bibfield{author}{\bibinfo{person}{Alison~L Calear} {and} \bibinfo{person}{Philip~J Batterham}.} \bibinfo{year}{2019}\natexlab{}.
\newblock \showarticletitle{Suicidal ideation disclosure: Patterns, correlates and outcome}.
\newblock \bibinfo{journal}{\emph{Psychiatry research}}  \bibinfo{volume}{278} (\bibinfo{year}{2019}), \bibinfo{pages}{1--6}.
\newblock


\bibitem[Chancellor and De~Choudhury(2020)]%
        {chancellor2020methods}
\bibfield{author}{\bibinfo{person}{Stevie Chancellor} {and} \bibinfo{person}{Munmun De~Choudhury}.} \bibinfo{year}{2020}\natexlab{}.
\newblock \showarticletitle{Methods in predictive techniques for mental health status on social media: a critical review}.
\newblock \bibinfo{journal}{\emph{NPJ digital medicine}} \bibinfo{volume}{3}, \bibinfo{number}{1} (\bibinfo{year}{2020}), \bibinfo{pages}{1--11}.
\newblock


\bibitem[Chen et~al\mbox{.}(2023)]%
        {chen2023two_stream_mdd}
\bibfield{author}{\bibinfo{person}{Siyuan Chen}, \bibinfo{person}{Zhiling Zhang}, \bibinfo{person}{Mengyue Wu}, {and} \bibinfo{person}{Kenny Zhu}.} \bibinfo{year}{2023}\natexlab{}.
\newblock \showarticletitle{Detection of multiple mental disorders from social media with two-stream psychiatric experts}. In \bibinfo{booktitle}{\emph{Proc. EMNLP}}.
\newblock


\bibitem[Clark et~al\mbox{.}(2020)]%
        {clark2020electra}
\bibfield{author}{\bibinfo{person}{Kevin Clark}, \bibinfo{person}{Minh-Thang Luong}, \bibinfo{person}{Quoc~V Le}, {and} \bibinfo{person}{Christopher~D Manning}.} \bibinfo{year}{2020}\natexlab{}.
\newblock \showarticletitle{Electra: Pre-training text encoders as discriminators rather than generators}.
\newblock \bibinfo{journal}{\emph{arXiv preprint arXiv:2003.10555}} (\bibinfo{year}{2020}).
\newblock


\bibitem[Copeland(2021)]%
        {copeland2021long}
\bibfield{author}{\bibinfo{person}{Molly Copeland}.} \bibinfo{year}{2021}\natexlab{}.
\newblock \showarticletitle{The long shadow of peers: adolescent networks and young adult mental health}.
\newblock \bibinfo{journal}{\emph{Social Sciences}} \bibinfo{volume}{10}, \bibinfo{number}{6} (\bibinfo{year}{2021}), \bibinfo{pages}{231}.
\newblock


\bibitem[Coppersmith et~al\mbox{.}(2014)]%
        {coppersmith2014quantifying}
\bibfield{author}{\bibinfo{person}{Glen Coppersmith}, \bibinfo{person}{Mark Dredze}, {and} \bibinfo{person}{Craig Harman}.} \bibinfo{year}{2014}\natexlab{}.
\newblock \showarticletitle{Quantifying mental health signals in Twitter}. In \bibinfo{booktitle}{\emph{Proceedings of the workshop on computational linguistics and clinical psychology: From linguistic signal to clinical reality}}. \bibinfo{pages}{51--60}.
\newblock


\bibitem[Coppersmith et~al\mbox{.}(2016)]%
        {coppersmith2016exploratory}
\bibfield{author}{\bibinfo{person}{Glen Coppersmith}, \bibinfo{person}{Kim Ngo}, \bibinfo{person}{Ryan Leary}, {and} \bibinfo{person}{Anthony Wood}.} \bibinfo{year}{2016}\natexlab{}.
\newblock \showarticletitle{Exploratory analysis of social media prior to a suicide attempt}. In \bibinfo{booktitle}{\emph{Proceedings of the third workshop on computational linguistics and clinical psychology}}. \bibinfo{pages}{106--117}.
\newblock


\bibitem[De~Choudhury and De(2014)]%
        {de2014mental}
\bibfield{author}{\bibinfo{person}{Munmun De~Choudhury} {and} \bibinfo{person}{Sushovan De}.} \bibinfo{year}{2014}\natexlab{}.
\newblock \showarticletitle{Mental health discourse on reddit: Self-disclosure, social support, and anonymity}. In \bibinfo{booktitle}{\emph{Proceedings of the international AAAI conference on web and social media}}, Vol.~\bibinfo{volume}{8}. \bibinfo{pages}{71--80}.
\newblock


\bibitem[De~Choudhury et~al\mbox{.}(2013)]%
        {dechoudhury2013predicting}
\bibfield{author}{\bibinfo{person}{Munmun De~Choudhury}, \bibinfo{person}{Michael Gamon}, \bibinfo{person}{Scott Counts}, {and} \bibinfo{person}{Eric Horvitz}.} \bibinfo{year}{2013}\natexlab{}.
\newblock \showarticletitle{Predicting depression via social media}. In \bibinfo{booktitle}{\emph{ICWSM}}.
\newblock


\bibitem[De~Choudhury and Kiciman(2017)]%
        {de2017language}
\bibfield{author}{\bibinfo{person}{Munmun De~Choudhury} {and} \bibinfo{person}{Emre Kiciman}.} \bibinfo{year}{2017}\natexlab{}.
\newblock \showarticletitle{The language of social support in social media and its effect on suicidal ideation risk}. In \bibinfo{booktitle}{\emph{ICWSM}}.
\newblock


\bibitem[De~Choudhury et~al\mbox{.}(2016a)]%
        {de2016discovering}
\bibfield{author}{\bibinfo{person}{Munmun De~Choudhury}, \bibinfo{person}{Emre Kiciman}, \bibinfo{person}{Mark Dredze}, \bibinfo{person}{Glen Coppersmith}, {and} \bibinfo{person}{Mrinal Kumar}.} \bibinfo{year}{2016}\natexlab{a}.
\newblock \showarticletitle{Discovering shifts to suicidal ideation from mental health content in social media}. In \bibinfo{booktitle}{\emph{CHI}}.
\newblock


\bibitem[De~Choudhury et~al\mbox{.}(2016b)]%
        {dechoudhury2016characterizing}
\bibfield{author}{\bibinfo{person}{Munmun De~Choudhury}, \bibinfo{person}{Emre Kiciman}, \bibinfo{person}{Mark Dredze}, \bibinfo{person}{Glen Coppersmith}, {and} \bibinfo{person}{Mrinal Kumar}.} \bibinfo{year}{2016}\natexlab{b}.
\newblock \showarticletitle{Discovering shifts to suicidal ideation from mental health content in social media}. In \bibinfo{booktitle}{\emph{Proceedings of the 2016 CHI conference on human factors in computing systems}}. \bibinfo{pages}{2098--2110}.
\newblock


\bibitem[Devlin et~al\mbox{.}(2018)]%
        {bert_base}
\bibfield{author}{\bibinfo{person}{Jacob Devlin}, \bibinfo{person}{Ming{-}Wei Chang}, \bibinfo{person}{Kenton Lee}, {and} \bibinfo{person}{Kristina Toutanova}.} \bibinfo{year}{2018}\natexlab{}.
\newblock \showarticletitle{{BERT:} Pre-training of Deep Bidirectional Transformers for Language Understanding}.
\newblock \bibinfo{journal}{\emph{CoRR}}  \bibinfo{volume}{abs/1810.04805} (\bibinfo{year}{2018}).
\newblock
\showeprint[arxiv]{1810.04805}
\urldef\tempurl%
\url{http://arxiv.org/abs/1810.04805}
\showURL{%
\tempurl}


\bibitem[Ernala et~al\mbox{.}(2017)]%
        {ernala2017linguistic}
\bibfield{author}{\bibinfo{person}{Sindhu~Kiranmai Ernala}, \bibinfo{person}{Asra~F Rizvi}, \bibinfo{person}{Michael~L Birnbaum}, \bibinfo{person}{John~M Kane}, {and} \bibinfo{person}{Munmun De~Choudhury}.} \bibinfo{year}{2017}\natexlab{}.
\newblock \showarticletitle{Linguistic markers indicating therapeutic outcomes of social media disclosures of schizophrenia}.
\newblock \bibinfo{journal}{\emph{Proceedings of the ACM on Human-Computer Interaction}} \bibinfo{volume}{1}, \bibinfo{number}{CSCW} (\bibinfo{year}{2017}), \bibinfo{pages}{1--27}.
\newblock


\bibitem[Fatima et~al\mbox{.}(2021)]%
        {fatima2021dasentimental}
\bibfield{author}{\bibinfo{person}{Asra Fatima}, \bibinfo{person}{Ying Li}, \bibinfo{person}{Thomas~Trenholm Hills}, {and} \bibinfo{person}{Massimo Stella}.} \bibinfo{year}{2021}\natexlab{}.
\newblock \showarticletitle{Dasentimental: Detecting depression, anxiety, and stress in texts via emotional recall, cognitive networks, and machine learning}.
\newblock \bibinfo{journal}{\emph{Big Data and Cognitive Computing}} \bibinfo{volume}{5}, \bibinfo{number}{4} (\bibinfo{year}{2021}), \bibinfo{pages}{77}.
\newblock


\bibitem[Franklin et~al\mbox{.}(2017)]%
        {franklin2017risk}
\bibfield{author}{\bibinfo{person}{Joseph~C Franklin}, \bibinfo{person}{Jessica~D Ribeiro}, \bibinfo{person}{Kathryn~R Fox}, \bibinfo{person}{Kate~H Bentley}, \bibinfo{person}{Evan~M Kleiman}, \bibinfo{person}{Xieyining Huang}, \bibinfo{person}{Katherine~M Musacchio}, \bibinfo{person}{Adam~C Jaroszewski}, \bibinfo{person}{Bernard~P Chang}, {and} \bibinfo{person}{Matthew~K Nock}.} \bibinfo{year}{2017}\natexlab{}.
\newblock \showarticletitle{Risk factors for suicidal thoughts and behaviors: A meta-analysis of 50 years of research.}
\newblock \bibinfo{journal}{\emph{Psychological bulletin}} (\bibinfo{year}{2017}).
\newblock


\bibitem[Gkotsis et~al\mbox{.}(2017)]%
        {gkotsis2017characterisation}
\bibfield{author}{\bibinfo{person}{George Gkotsis}, \bibinfo{person}{Anika Oellrich}, \bibinfo{person}{Sumithra Velupillai}, \bibinfo{person}{Maria Liakata}, \bibinfo{person}{Tim~JP Hubbard}, \bibinfo{person}{Richard~JB Dobson}, {and} \bibinfo{person}{Rina Dutta}.} \bibinfo{year}{2017}\natexlab{}.
\newblock \showarticletitle{Characterisation of mental health conditions in social media using Informed Deep Learning}.
\newblock \bibinfo{journal}{\emph{Scientific reports}} (\bibinfo{year}{2017}).
\newblock


\bibitem[Goyal et~al\mbox{.}(2024)]%
        {goyal2024using}
\bibfield{author}{\bibinfo{person}{Abhay Goyal}, \bibinfo{person}{Roger Ho~Chun~Man}, \bibinfo{person}{Roy Ka-Wei Lee}, \bibinfo{person}{Koustuv Saha}, \bibinfo{person}{Frederick L.~Altice}, \bibinfo{person}{Christian Poellabauer}, \bibinfo{person}{Orestis Papakyriakopoulos}, \bibinfo{person}{Lam Yin~Cheung}, \bibinfo{person}{Munmun De~Choudhury}, \bibinfo{person}{Kanica Allagh}, {et~al\mbox{.}}} \bibinfo{year}{2024}\natexlab{}.
\newblock \showarticletitle{Using Voice Data to Facilitate Depression Risk Assessment in Primary Health Care}. In \bibinfo{booktitle}{\emph{Companion Publication of the 16th ACM Web Science Conference}}. \bibinfo{pages}{17--18}.
\newblock


\bibitem[Grootendorst(2022)]%
        {grootendorst2022bertopic}
\bibfield{author}{\bibinfo{person}{Maarten Grootendorst}.} \bibinfo{year}{2022}\natexlab{}.
\newblock \showarticletitle{BERTopic: Neural topic modeling with a class-based TF-IDF procedure}.
\newblock \bibinfo{journal}{\emph{arXiv preprint arXiv:2203.05794}} (\bibinfo{year}{2022}).
\newblock


\bibitem[Guntuku et~al\mbox{.}(2017)]%
        {guntuku2017detecting}
\bibfield{author}{\bibinfo{person}{Sharath~Chandra Guntuku}, \bibinfo{person}{David~B Yaden}, \bibinfo{person}{Margaret~L Kern}, \bibinfo{person}{Lyle~H Ungar}, {and} \bibinfo{person}{Johannes~C Eichstaedt}.} \bibinfo{year}{2017}\natexlab{}.
\newblock \showarticletitle{Detecting depression and mental illness on social media: an integrative review}.
\newblock \bibinfo{journal}{\emph{Current Opinion in Behavioral Sciences}}  \bibinfo{volume}{18} (\bibinfo{year}{2017}), \bibinfo{pages}{43--49}.
\newblock


\bibitem[Hallford et~al\mbox{.}(2023)]%
        {hallford2023disclosure}
\bibfield{author}{\bibinfo{person}{David~John Hallford}, \bibinfo{person}{Danielle Rusanov}, \bibinfo{person}{B Winestone}, \bibinfo{person}{R Kaplan}, \bibinfo{person}{Matthew Fuller-Tyszkiewicz}, {and} \bibinfo{person}{Glenn Melvin}.} \bibinfo{year}{2023}\natexlab{}.
\newblock \showarticletitle{Disclosure of suicidal ideation and behaviours: A systematic review and meta-analysis of prevalence}.
\newblock \bibinfo{journal}{\emph{Clinical psychology review}}  \bibinfo{volume}{101} (\bibinfo{year}{2023}), \bibinfo{pages}{102272}.
\newblock


\bibitem[He et~al\mbox{.}(2021)]%
        {he2021debertav3}
\bibfield{author}{\bibinfo{person}{Pengcheng He}, \bibinfo{person}{Jianfeng Gao}, {and} \bibinfo{person}{Weizhu Chen}.} \bibinfo{year}{2021}\natexlab{}.
\newblock \showarticletitle{Debertav3: Improving deberta using electra-style pre-training with gradient-disentangled embedding sharing}.
\newblock \bibinfo{journal}{\emph{arXiv preprint arXiv:2111.09543}} (\bibinfo{year}{2021}).
\newblock


\bibitem[Ji(2022)]%
        {ji2022towards}
\bibfield{author}{\bibinfo{person}{Shaoxiong Ji}.} \bibinfo{year}{2022}\natexlab{}.
\newblock \showarticletitle{Towards intention understanding in suicidal risk assessment with natural language processing}. In \bibinfo{booktitle}{\emph{EMNLP}}. The Association for Computational Linguistics.
\newblock


\bibitem[Johnson et~al\mbox{.}({[n.\,d.]})]%
        {johnson2022s}
\bibfield{author}{\bibinfo{person}{Jazette Johnson}, \bibinfo{person}{Vitica Arnold}, \bibinfo{person}{Anne~Marie Piper}, {and} \bibinfo{person}{Gillian~R Hayes}.} \bibinfo{year}{[n.\,d.]}\natexlab{}.
\newblock \showarticletitle{" It's a lonely disease": Cultivating Online Spaces for Social Support among People Living with Dementia and Dementia Caregivers}.
\newblock \bibinfo{journal}{\emph{PACM HCI}} \bibinfo{number}{CSCW2} (\bibinfo{year}{[n.\,d.]}).
\newblock


\bibitem[Joiner(2005)]%
        {joiner2005interpersonal}
\bibfield{author}{\bibinfo{person}{Thomas~E. Joiner}.} \bibinfo{year}{2005}\natexlab{}.
\newblock \bibinfo{booktitle}{\emph{Why People Die by Suicide}}.
\newblock \bibinfo{publisher}{Harvard University Press}.
\newblock


\bibitem[Kim et~al\mbox{.}(2023)]%
        {kim2023supporters}
\bibfield{author}{\bibinfo{person}{Meeyun Kim}, \bibinfo{person}{Koustuv Saha}, \bibinfo{person}{Munmun De~Choudhury}, {and} \bibinfo{person}{Daejin Choi}.} \bibinfo{year}{2023}\natexlab{}.
\newblock \showarticletitle{Supporters first: understanding online social support on mental health from a supporter perspective}.
\newblock \bibinfo{journal}{\emph{Proceedings of the ACM on Human-Computer Interaction}} \bibinfo{volume}{7}, \bibinfo{number}{CSCW1} (\bibinfo{year}{2023}), \bibinfo{pages}{1--28}.
\newblock


\bibitem[Laurer et~al\mbox{.}(2023)]%
        {laurer_less_2023}
\bibfield{author}{\bibinfo{person}{Moritz Laurer}, \bibinfo{person}{Wouter van Atteveldt}, \bibinfo{person}{Andreu~Salleras Casas}, {and} \bibinfo{person}{Kasper Welbers}.} \bibinfo{year}{2023}\natexlab{}.
\newblock \showarticletitle{Less Annotating, More Classifying: Addressing the Data Scarcity Issue of Supervised Machine Learning with Deep Transfer Learning and BERT-NLI}.
\newblock \bibinfo{journal}{\emph{Political Analysis}} (\bibinfo{date}{June} \bibinfo{year}{2023}), \bibinfo{pages}{1--33}.
\newblock
\showISSN{1047-1987, 1476-4989}
\href{https://doi.org/10.1017/pan.2023.20}{doi:\nolinkurl{10.1017/pan.2023.20}}


\bibitem[Liu et~al\mbox{.}(2019)]%
        {liu2019roberta}
\bibfield{author}{\bibinfo{person}{Yinhan Liu}, \bibinfo{person}{Myle Ott}, \bibinfo{person}{Naman Goyal}, \bibinfo{person}{Jingfei Du}, \bibinfo{person}{Mandar Joshi}, \bibinfo{person}{Danqi Chen}, \bibinfo{person}{Omer Levy}, \bibinfo{person}{Mike Lewis}, \bibinfo{person}{Luke Zettlemoyer}, {and} \bibinfo{person}{Veselin Stoyanov}.} \bibinfo{year}{2019}\natexlab{}.
\newblock \showarticletitle{Roberta: A robustly optimized bert pretraining approach}.
\newblock \bibinfo{journal}{\emph{arXiv:1907.11692}} (\bibinfo{year}{2019}).
\newblock


\bibitem[McGillivray et~al\mbox{.}(2022)]%
        {mcgillivray2022non}
\bibfield{author}{\bibinfo{person}{Lauren McGillivray}, \bibinfo{person}{Demee Rheinberger}, \bibinfo{person}{Jessica Wang}, \bibinfo{person}{Alexander Burnett}, {and} \bibinfo{person}{Michelle Torok}.} \bibinfo{year}{2022}\natexlab{}.
\newblock \showarticletitle{Non-disclosing youth: a cross sectional study to understand why young people do not disclose suicidal thoughts to their mental health professional}.
\newblock \bibinfo{journal}{\emph{BMC psychiatry}} \bibinfo{volume}{22}, \bibinfo{number}{1} (\bibinfo{year}{2022}), \bibinfo{pages}{3}.
\newblock


\bibitem[Nesi et~al\mbox{.}(2021)]%
        {nesi2021social}
\bibfield{author}{\bibinfo{person}{Jacqueline Nesi}, \bibinfo{person}{Taylor~A Burke}, \bibinfo{person}{Alexandra~H Bettis}, \bibinfo{person}{Anastacia~Y Kudinova}, \bibinfo{person}{Elizabeth~C Thompson}, \bibinfo{person}{Heather~A MacPherson}, \bibinfo{person}{Kara~A Fox}, \bibinfo{person}{Hannah~R Lawrence}, \bibinfo{person}{Sarah~A Thomas}, \bibinfo{person}{Jennifer~C Wolff}, {et~al\mbox{.}}} \bibinfo{year}{2021}\natexlab{}.
\newblock \showarticletitle{Social media use and self-injurious thoughts and behaviors: A systematic review and meta-analysis}.
\newblock \bibinfo{journal}{\emph{Clinical psychology review}}  \bibinfo{volume}{87} (\bibinfo{year}{2021}), \bibinfo{pages}{102038}.
\newblock


\bibitem[Niederkrotenthaler et~al\mbox{.}(2010)]%
        {niederkrotenthaler2010role}
\bibfield{author}{\bibinfo{person}{Thomas Niederkrotenthaler}, \bibinfo{person}{Martin Voracek}, \bibinfo{person}{Arno Herberth}, \bibinfo{person}{Benedikt Till}, \bibinfo{person}{Markus Strauss}, \bibinfo{person}{Elmar Etzersdorfer}, \bibinfo{person}{Brigitte Eisenwort}, {and} \bibinfo{person}{Gernot Sonneck}.} \bibinfo{year}{2010}\natexlab{}.
\newblock \showarticletitle{Role of media reports in completed and prevented suicide: Werther v. Papageno effects}.
\newblock \bibinfo{journal}{\emph{The British Journal of Psychiatry}} \bibinfo{volume}{197}, \bibinfo{number}{3} (\bibinfo{year}{2010}), \bibinfo{pages}{234--243}.
\newblock


\bibitem[O'Connor and Kirtley(2018)]%
        {o2018integrated}
\bibfield{author}{\bibinfo{person}{Rory~C O'Connor} {and} \bibinfo{person}{Olivia~J Kirtley}.} \bibinfo{year}{2018}\natexlab{}.
\newblock \showarticletitle{The integrated motivational--volitional model of suicidal behaviour}.
\newblock \bibinfo{journal}{\emph{Philosophical Transactions of the Royal Society B: Biological Sciences}} \bibinfo{volume}{373}, \bibinfo{number}{1754} (\bibinfo{year}{2018}), \bibinfo{pages}{20170268}.
\newblock


\bibitem[Pennebaker and Chung(2007)]%
        {pennebaker2007expressive}
\bibfield{author}{\bibinfo{person}{James~W Pennebaker} {and} \bibinfo{person}{Cindy~K Chung}.} \bibinfo{year}{2007}\natexlab{}.
\newblock \showarticletitle{Expressive writing, emotional upheavals, and health}.
\newblock \bibinfo{journal}{\emph{Handbook of health psychology}} (\bibinfo{year}{2007}), \bibinfo{pages}{263--284}.
\newblock


\bibitem[P{\'e}rez et~al\mbox{.}(2022)]%
        {perez2023semantic_severity}
\bibfield{author}{\bibinfo{person}{Anxo P{\'e}rez}, \bibinfo{person}{Neha Warikoo}, \bibinfo{person}{Kexin Wang}, \bibinfo{person}{Javier Parapar}, {and} \bibinfo{person}{Iryna Gurevych}.} \bibinfo{year}{2022}\natexlab{}.
\newblock \showarticletitle{Semantic similarity models for depression severity estimation}.
\newblock \bibinfo{journal}{\emph{arXiv preprint arXiv:2211.07624}} (\bibinfo{year}{2022}).
\newblock


\bibitem[Phillips(1974)]%
        {phillips1974influence}
\bibfield{author}{\bibinfo{person}{David~P Phillips}.} \bibinfo{year}{1974}\natexlab{}.
\newblock \showarticletitle{The influence of suggestion on suicide: Substantive and theoretical implications of the Werther effect}.
\newblock \bibinfo{journal}{\emph{American sociological review}} (\bibinfo{year}{1974}).
\newblock


\bibitem[Podlogar et~al\mbox{.}(2022)]%
        {podlogar2022past}
\bibfield{author}{\bibinfo{person}{Matthew~C Podlogar}, \bibinfo{person}{Peter~M Gutierrez}, {and} \bibinfo{person}{Thomas~E Joiner}.} \bibinfo{year}{2022}\natexlab{}.
\newblock \showarticletitle{Past levels of mental health intervention and current nondisclosure of suicide risk among men older than age 50}.
\newblock \bibinfo{journal}{\emph{Assessment}} \bibinfo{volume}{29}, \bibinfo{number}{8} (\bibinfo{year}{2022}), \bibinfo{pages}{1611--1621}.
\newblock


\bibitem[Qiu et~al\mbox{.}(2024)]%
        {qiu-etal-2024-psyguard}
\bibfield{author}{\bibinfo{person}{Huachuan Qiu}, \bibinfo{person}{Lizhi Ma}, {and} \bibinfo{person}{Zhenzhong Lan}.} \bibinfo{year}{2024}\natexlab{}.
\newblock \showarticletitle{PsyGUARD: An automated system for suicide detection and risk assessment in psychological counseling}.
\newblock \bibinfo{journal}{\emph{arXiv preprint arXiv:2409.20243}} (\bibinfo{year}{2024}).
\newblock


\bibitem[Rawat et~al\mbox{.}(2022)]%
        {rawat2022scan}
\bibfield{author}{\bibinfo{person}{Bhanu Pratap~Singh Rawat}, \bibinfo{person}{Samuel Kovaly}, \bibinfo{person}{Wilfred~R Pigeon}, {and} \bibinfo{person}{Hong Yu}.} \bibinfo{year}{2022}\natexlab{}.
\newblock \showarticletitle{Scan: suicide attempt and ideation events dataset}. In \bibinfo{booktitle}{\emph{NAACL}}.
\newblock


\bibitem[Saha et~al\mbox{.}(2020a)]%
        {saha2020omhc}
\bibfield{author}{\bibinfo{person}{Koustuv Saha}, \bibinfo{person}{Sindhu~Kiranmai Ernala}, \bibinfo{person}{Sarmistha Dutta}, \bibinfo{person}{Eva Sharma}, {and} \bibinfo{person}{Munmun De~Choudhury}.} \bibinfo{year}{2020}\natexlab{a}.
\newblock \showarticletitle{Understanding Moderation in Online Mental Health Communities}. In \bibinfo{booktitle}{\emph{HCII}}. Springer.
\newblock


\bibitem[Saha et~al\mbox{.}(2021)]%
        {saha2021person}
\bibfield{author}{\bibinfo{person}{Koustuv Saha}, \bibinfo{person}{Ted Grover}, \bibinfo{person}{Stephen~M Mattingly}, \bibinfo{person}{Vedant~Das swain}, \bibinfo{person}{Pranshu Gupta}, \bibinfo{person}{Gonzalo~J Martinez}, \bibinfo{person}{Pablo Robles-Granda}, \bibinfo{person}{Gloria Mark}, \bibinfo{person}{Aaron Striegel}, {and} \bibinfo{person}{Munmun De~Choudhury}.} \bibinfo{year}{2021}\natexlab{}.
\newblock \showarticletitle{Person-Centered Predictions of Psychological Constructs with Social Media Contextualized by Multimodal Sensing}.
\newblock \bibinfo{journal}{\emph{PACM IMWUT}} (\bibinfo{year}{2021}).
\newblock


\bibitem[Saha et~al\mbox{.}(2024)]%
        {saha2024observer}
\bibfield{author}{\bibinfo{person}{Koustuv Saha}, \bibinfo{person}{Pranshu Gupta}, \bibinfo{person}{Gloria Mark}, \bibinfo{person}{Emre Kiciman}, {and} \bibinfo{person}{Munmun De~Choudhury}.} \bibinfo{year}{2024}\natexlab{}.
\newblock \showarticletitle{Observer Effect in Social Media Use}. In \bibinfo{booktitle}{\emph{Proceedings of the 2024 CHI Conference on Human Factors in Computing Systems}}.
\newblock


\bibitem[Saha et~al\mbox{.}(2025)]%
        {saha2025mental}
\bibfield{author}{\bibinfo{person}{Koustuv Saha}, \bibinfo{person}{Bhaskar Kotakonda}, {and} \bibinfo{person}{Munmun De~Choudhury}.} \bibinfo{year}{2025}\natexlab{}.
\newblock \showarticletitle{Mental Health Impact of the COVID-19 Pandemic on College Students: A Quasi-Experimental Study on Social Media}. In \bibinfo{booktitle}{\emph{Proceedings of the International AAAI Conference on Web and Social Media}}.
\newblock


\bibitem[Saha and Sharma(2020)]%
        {saha2020causal}
\bibfield{author}{\bibinfo{person}{Koustuv Saha} {and} \bibinfo{person}{Amit Sharma}.} \bibinfo{year}{2020}\natexlab{}.
\newblock \showarticletitle{Causal factors of effective psychosocial outcomes in online mental health communities}. In \bibinfo{booktitle}{\emph{ICWSM}}.
\newblock


\bibitem[Saha et~al\mbox{.}(2019)]%
        {saha2019social}
\bibfield{author}{\bibinfo{person}{Koustuv Saha}, \bibinfo{person}{Benjamin Sugar}, \bibinfo{person}{John Torous}, \bibinfo{person}{Bruno Abrahao}, \bibinfo{person}{Emre K{\i}c{\i}man}, {and} \bibinfo{person}{Munmun De~Choudhury}.} \bibinfo{year}{2019}\natexlab{}.
\newblock \showarticletitle{A social media study on the effects of psychiatric medication use}. In \bibinfo{booktitle}{\emph{Proceedings of the International AAAI Conference on Web and Social Media}}.
\newblock


\bibitem[Saha et~al\mbox{.}(2020b)]%
        {saha2020psychosocial}
\bibfield{author}{\bibinfo{person}{Koustuv Saha}, \bibinfo{person}{John Torous}, \bibinfo{person}{Eric~D Caine}, {and} \bibinfo{person}{Munmun De~Choudhury}.} \bibinfo{year}{2020}\natexlab{b}.
\newblock \showarticletitle{Psychosocial Effects of the COVID-19 Pandemic: Large-scale Quasi-Experimental Study on Social Media}.
\newblock \bibinfo{journal}{\emph{Journal of medical internet research}} \bibinfo{volume}{22}, \bibinfo{number}{11} (\bibinfo{year}{2020}), \bibinfo{pages}{e22600}.
\newblock


\bibitem[Saha et~al\mbox{.}(2022)]%
        {saha2022social}
\bibfield{author}{\bibinfo{person}{Koustuv Saha}, \bibinfo{person}{Asra Yousuf}, \bibinfo{person}{Ryan~L Boyd}, \bibinfo{person}{James~W Pennebaker}, {and} \bibinfo{person}{Munmun De~Choudhury}.} \bibinfo{year}{2022}\natexlab{}.
\newblock \showarticletitle{Social media discussions predict mental health consultations on college campuses}.
\newblock \bibinfo{journal}{\emph{Scientific reports}} \bibinfo{volume}{12}, \bibinfo{number}{1} (\bibinfo{year}{2022}), \bibinfo{pages}{123}.
\newblock


\bibitem[Sawhney et~al\mbox{.}(2021a)]%
        {sawhney2021phase}
\bibfield{author}{\bibinfo{person}{Ramit Sawhney}, \bibinfo{person}{Harshit Joshi}, \bibinfo{person}{Lucie Flek}, {and} \bibinfo{person}{Rajiv Shah}.} \bibinfo{year}{2021}\natexlab{a}.
\newblock \showarticletitle{Phase: Learning emotional phase-aware representations for suicide ideation detection on social media}. In \bibinfo{booktitle}{\emph{Proceedings of the 16th conference of the European Chapter of the Association for Computational Linguistics: main volume}}. \bibinfo{pages}{2415--2428}.
\newblock


\bibitem[Sawhney et~al\mbox{.}(2020)]%
        {sawhney2020statenet}
\bibfield{author}{\bibinfo{person}{Ramit Sawhney}, \bibinfo{person}{Harshit Joshi}, \bibinfo{person}{Saumya Gandhi}, {and} \bibinfo{person}{Rajiv Shah}.} \bibinfo{year}{2020}\natexlab{}.
\newblock \showarticletitle{A time-aware transformer based model for suicide ideation detection on social media}. In \bibinfo{booktitle}{\emph{EMNLP}}.
\newblock


\bibitem[Sawhney et~al\mbox{.}(2021b)]%
        {sawhney2021hyperbolic}
\bibfield{author}{\bibinfo{person}{Ramit Sawhney}, \bibinfo{person}{Harshit Joshi}, \bibinfo{person}{Rajiv Shah}, {and} \bibinfo{person}{Lucie Flek}.} \bibinfo{year}{2021}\natexlab{b}.
\newblock \showarticletitle{Suicide ideation detection via social and temporal user representations using hyperbolic learning}. In \bibinfo{booktitle}{\emph{NAACL}}.
\newblock


\bibitem[Sharma et~al\mbox{.}(2020)]%
        {sharma2020computational}
\bibfield{author}{\bibinfo{person}{Ashish Sharma}, \bibinfo{person}{Adam Miner}, \bibinfo{person}{David Atkins}, {and} \bibinfo{person}{Tim Althoff}.} \bibinfo{year}{2020}\natexlab{}.
\newblock \showarticletitle{A Computational Approach to Understanding Empathy Expressed in Text-Based Mental Health Support}. In \bibinfo{booktitle}{\emph{EMNLP}}.
\newblock


\bibitem[Sharma and De~Choudhury(2018)]%
        {sharma2018mental}
\bibfield{author}{\bibinfo{person}{Eva Sharma} {and} \bibinfo{person}{Munmun De~Choudhury}.} \bibinfo{year}{2018}\natexlab{}.
\newblock \showarticletitle{Mental health support and its relationship to linguistic accommodation in online communities}. In \bibinfo{booktitle}{\emph{Proceedings of the 2018 CHI conference on human factors in computing systems}}. \bibinfo{pages}{1--13}.
\newblock


\bibitem[Shen et~al\mbox{.}(2020)]%
        {shen2020suicide}
\bibfield{author}{\bibinfo{person}{Chen Shen} {et~al\mbox{.}}} \bibinfo{year}{2020}\natexlab{}.
\newblock \showarticletitle{Suicide risk prediction using social media data}. In \bibinfo{booktitle}{\emph{AAAI}}.
\newblock


\bibitem[Shimgekar et~al\mbox{.}(2025)]%
        {shimgekar2025interpersonal}
\bibfield{author}{\bibinfo{person}{Soorya~Ram Shimgekar}, \bibinfo{person}{Violeta~J Rodriguez}, \bibinfo{person}{Paul~A Bloom}, \bibinfo{person}{Dong~Whi Yoo}, {and} \bibinfo{person}{Koustuv Saha}.} \bibinfo{year}{2025}\natexlab{}.
\newblock \showarticletitle{Interpersonal Theory of Suicide as a Lens to Examine Suicidal Ideation in Online Spaces}.
\newblock \bibinfo{journal}{\emph{arXiv preprint arXiv:2504.13277}} (\bibinfo{year}{2025}).
\newblock


\bibitem[Shin et~al\mbox{.}(2024)]%
        {shin2024using}
\bibfield{author}{\bibinfo{person}{Daun Shin}, \bibinfo{person}{Hyoseung Kim}, \bibinfo{person}{Seunghwan Lee}, \bibinfo{person}{Younhee Cho}, {and} \bibinfo{person}{Whanbo Jung}.} \bibinfo{year}{2024}\natexlab{}.
\newblock \showarticletitle{Using large language models to detect depression from user-generated diary text data as a novel approach in digital mental health screening: instrument validation study}.
\newblock \bibinfo{journal}{\emph{Journal of Medical Internet Research}}  \bibinfo{volume}{26} (\bibinfo{year}{2024}), \bibinfo{pages}{e54617}.
\newblock


\bibitem[Tausczik and Pennebaker(2010)]%
        {tausczik2010psychological}
\bibfield{author}{\bibinfo{person}{Yla~R Tausczik} {and} \bibinfo{person}{James~W Pennebaker}.} \bibinfo{year}{2010}\natexlab{}.
\newblock \showarticletitle{The psychological meaning of words: LIWC and computerized text analysis methods}.
\newblock \bibinfo{journal}{\emph{Journal of language and social psychology}} \bibinfo{volume}{29}, \bibinfo{number}{1} (\bibinfo{year}{2010}), \bibinfo{pages}{24--54}.
\newblock


\bibitem[Teng et~al\mbox{.}(2025)]%
        {teng2025enhancing}
\bibfield{author}{\bibinfo{person}{Shiyu Teng}, \bibinfo{person}{Jiaqing Liu}, \bibinfo{person}{Rahul~Kumar Jain}, \bibinfo{person}{Shurong Chai}, \bibinfo{person}{Ruibo Hou}, \bibinfo{person}{Tomoko Tateyama}, \bibinfo{person}{Lanfen Lin}, {and} \bibinfo{person}{Yen-wei Chen}.} \bibinfo{year}{2025}\natexlab{}.
\newblock \showarticletitle{Enhancing Depression Detection with Chain-of-Thought Prompting: From Emotion to Reasoning Using Large Language Models}.
\newblock \bibinfo{journal}{\emph{arXiv preprint arXiv:2502.05879}} (\bibinfo{year}{2025}).
\newblock


\bibitem[Thieme et~al\mbox{.}(2020)]%
        {thieme2020machine}
\bibfield{author}{\bibinfo{person}{Anja Thieme}, \bibinfo{person}{Danielle Belgrave}, {and} \bibinfo{person}{Gavin Doherty}.} \bibinfo{year}{2020}\natexlab{}.
\newblock \showarticletitle{Machine learning in mental health: A systematic review of the HCI literature to support the development of effective and implementable ML systems}.
\newblock \bibinfo{journal}{\emph{ACM Transactions on Computer-Human Interaction}} (\bibinfo{year}{2020}).
\newblock


\bibitem[Trotzek et~al\mbox{.}(2018)]%
        {trotzek2018utilizing}
\bibfield{author}{\bibinfo{person}{Marcel Trotzek}, \bibinfo{person}{Sven Koitka}, {and} \bibinfo{person}{Christoph~M Friedrich}.} \bibinfo{year}{2018}\natexlab{}.
\newblock \showarticletitle{Utilizing neural networks and linguistic metadata for early detection of depression indications in text sequences}.
\newblock \bibinfo{journal}{\emph{IEEE Transactions on Knowledge and Data Engineering}} \bibinfo{volume}{32}, \bibinfo{number}{3} (\bibinfo{year}{2018}), \bibinfo{pages}{588--601}.
\newblock


\bibitem[Wang et~al\mbox{.}(2024)]%
        {wang2024_explainable_depression}
\bibfield{author}{\bibinfo{person}{Yuxi Wang}, \bibinfo{person}{Diana Inkpen}, {and} \bibinfo{person}{Prasadith~Kirinde Gamaarachchige}.} \bibinfo{year}{2024}\natexlab{}.
\newblock \showarticletitle{Explainable depression detection using large language models on social media data}. In \bibinfo{booktitle}{\emph{Proceedings of the 9th Workshop on Computational Linguistics and Clinical Psychology (CLPsych 2024)}}. \bibinfo{pages}{108--126}.
\newblock


\bibitem[WHO(2025)]%
        {world2025suicide}
\bibfield{author}{\bibinfo{person}{WHO}.} \bibinfo{year}{2025}\natexlab{}.
\newblock \bibinfo{booktitle}{\emph{Suicide}}.
\newblock \bibinfo{publisher}{World Health Organization}.
\newblock
\newblock
\shownote{https://www.who.int/news-room/fact-sheets/detail/suicide}.


\bibitem[Wyman et~al\mbox{.}(2019)]%
        {wyman2019peer}
\bibfield{author}{\bibinfo{person}{Peter~A Wyman}, \bibinfo{person}{Trevor~A Pickering}, \bibinfo{person}{Anthony~R Pisani}, \bibinfo{person}{Kelly Rulison}, \bibinfo{person}{Karen Schmeelk-Cone}, \bibinfo{person}{Chelsey Hartley}, \bibinfo{person}{Madelyn Gould}, \bibinfo{person}{Eric~D Caine}, \bibinfo{person}{Mark LoMurray}, \bibinfo{person}{Charles~Hendricks Brown}, {et~al\mbox{.}}} \bibinfo{year}{2019}\natexlab{}.
\newblock \showarticletitle{Peer-adult network structure and suicide attempts in 38 high schools: Implications for network-informed suicide prevention}.
\newblock \bibinfo{journal}{\emph{Journal of Child Psychology and Psychiatry}} \bibinfo{volume}{60}, \bibinfo{number}{10} (\bibinfo{year}{2019}), \bibinfo{pages}{1065--1075}.
\newblock


\bibitem[Yuan et~al\mbox{.}(2023)]%
        {yuan2023mental}
\bibfield{author}{\bibinfo{person}{Yunhao Yuan}, \bibinfo{person}{Koustuv Saha}, \bibinfo{person}{Barbara Keller}, \bibinfo{person}{Erkki~Tapio Isomets{\"a}}, {and} \bibinfo{person}{Talayeh Aledavood}.} \bibinfo{year}{2023}\natexlab{}.
\newblock \showarticletitle{Mental Health Coping Stories on Social Media: A Causal-Inference Study of Papageno Effect}. In \bibinfo{booktitle}{\emph{Proceedings of the ACM Web Conference 2023}}. \bibinfo{pages}{2677--2685}.
\newblock


\bibitem[Zhang et~al\mbox{.}(2025)]%
        {zhang2025ketch}
\bibfield{author}{\bibinfo{person}{Dongsong Zhang}, \bibinfo{person}{Lina Zhou}, \bibinfo{person}{Jie Tao}, \bibinfo{person}{Tingshao Zhu}, {and} \bibinfo{person}{Guodong Gao}.} \bibinfo{year}{2025}\natexlab{}.
\newblock \showarticletitle{KETCH: A knowledge-enhanced transformer-based approach to suicidal ideation detection from social media content}.
\newblock \bibinfo{journal}{\emph{Information Systems Research}} \bibinfo{volume}{36}, \bibinfo{number}{1} (\bibinfo{year}{2025}).
\newblock


\bibitem[Zhang et~al\mbox{.}(2022)]%
        {zhang2022psysym}
\bibfield{author}{\bibinfo{person}{Zhiling Zhang}, \bibinfo{person}{Siyuan Chen}, \bibinfo{person}{Mengyue Wu}, {and} \bibinfo{person}{Kenny~Q Zhu}.} \bibinfo{year}{2022}\natexlab{}.
\newblock \showarticletitle{Symptom identification for interpretable detection of multiple mental disorders}.
\newblock \bibinfo{journal}{\emph{arXiv preprint arXiv:2205.11308}} (\bibinfo{year}{2022}).
\newblock


\bibitem[Zhou et~al\mbox{.}(2022)]%
        {zhou2022veteran}
\bibfield{author}{\bibinfo{person}{Jiawei Zhou}, \bibinfo{person}{Koustuv Saha}, \bibinfo{person}{Irene~Michelle Lopez~Carron}, \bibinfo{person}{Dong~Whi Yoo}, \bibinfo{person}{Catherine~R Deeter}, \bibinfo{person}{Munmun De~Choudhury}, {and} \bibinfo{person}{Rosa~I Arriaga}.} \bibinfo{year}{2022}\natexlab{}.
\newblock \showarticletitle{Veteran Critical Theory as a Lens to Understand Veterans' Needs and Support on Social Media}.
\newblock \bibinfo{journal}{\emph{Proceedings of the ACM on Human-Computer Interaction}} \bibinfo{volume}{6}, \bibinfo{number}{CSCW1} (\bibinfo{year}{2022}), \bibinfo{pages}{1--28}.
\newblock


\bibitem[Zirikly et~al\mbox{.}(2019)]%
        {zirikly2019clpsych}
\bibfield{author}{\bibinfo{person}{Ayah Zirikly}, \bibinfo{person}{Philip Resnik}, \bibinfo{person}{Ozlem Uzuner}, {and} \bibinfo{person}{Kristy Hollingshead}.} \bibinfo{year}{2019}\natexlab{}.
\newblock \showarticletitle{CLPsych 2019 shared task: Predicting the degree of suicide risk in Reddit posts}. In \bibinfo{booktitle}{\emph{Proceedings of the sixth workshop on computational linguistics and clinical psychology}}. \bibinfo{pages}{24--33}.
\newblock


\end{thebibliography}
